\documentstyle[12pt,epsf]{article}
\newcommand{\nsigma}{\mbox{\boldmath $\sigma$}}

\newcommand{\nq}{{\bf q}}
\newcommand{\np}{{\bf p}}

\newcommand{\nk}{{\bf k}}
\newcommand{\nr}{{\bf r}}

\newcommand{\e}[1]{ {\rm e}^{#1} }
\newcommand{\half}{\frac{1}{2}}
 
\newcommand{\sixj}[6]{ \left\{ \begin{array}{ccc}
                               #1 & #2 & #3 \\
                                #4 & #5 & #6 
                               \end{array}
                        \right\} } 
\begin{document}
\vspace*{1cm}
\centerline{\large {\bf Mean field description of electron induced}}
\centerline{\large {\bf   quasi-elastic excitation in nuclei} }
\vspace{1.5cm}
\centerline{\bf J. Enrique Amaro}
\centerline{Departamento de F\'{\i}sica Moderna, Universidad de Granada,}
\centerline{E-18071 Granada, Spain}

\vspace{.5cm}

\centerline{\bf Giampaolo Co'}
\centerline{Dipartimento di Fisica, Universit\`a di Lecce}
\centerline{and I.N.F.N. sezione di Lecce, I-73100 Lecce, Italy}

\vspace{.5cm}

\centerline{\bf Antonio M. Lallena}
\centerline{Departamento de F\'{\i}sica Moderna, Universidad de Granada,}
\centerline{E-18071 Granada, Spain}

\vspace{1.5cm}

\begin{abstract}
The validity of the approximations done in the mean field description
of the quasi-elastic excitation of medium-heavy nuclei is discussed.
A test of the reliability of the plane wave Born approximation is
presented. The uncertainty related to the choice of the
electromagnetic nucleon form factors is discussed. The effects
produced by the meson exchange currents generated by the exchange of a
single pion are studied, as well as the need of including relativistic
corrections to the one-body currents. The results of the continuum
shell model and those of the Fermi gas model are compared and the
need of treating the emitted nucleon within a relativistic framework
is studied. We analyze the results of Random Phase Approximation
calculations not only in terms of their effects on the mean field
responses, but also in terms of the theoretical consistency in the
choice of the effective residual interaction. The role of the final
state interaction is investigated. A microscopic justification of the
use of complex optical potential to describe the final state
interaction is provided. The effects of the final state interaction
on the mean field responses are studied with a sum rule conserving
model. A comparison  with the experimental data measured in the
$^{12}$C and $^{40}$Ca nuclei is shown.
\end{abstract}

\newpage
\section{Introduction}

The Mean Field (MF) model is the basis of any description of
many-particle systems. In this model the particles composing the
system are supposed to move independently from each other
in an average potential. 
Under this assumption the complicated many-body problem is
transformed into a set of, easy to solve, one-body problems. 
The solution
of the MF equations produces a set of orthonormal single particle (sp)
states whose tensor product
forms a basis in the many-body Hilbert space. This basis is
used in more complicated  many-body treatments. 

Usually the behavior of the real many-particle systems is quite
different from the MF predictions. There are however phenomena which
can be rather well described by the MF model. In these cases the more
complicated many-body effects can be treated as a correction to the MF
solution. One of these phenomena is the quasi-elastic (QE) excitation
which is dominated by the single particle dynamics, well described by
the MF approach in terms of one-particle one-hole (1p-1h) excitations.

Experimentally, the QE excitation is obtained by using probes which
interact weakly with the system and by controlling the kinematical
conditions in such a way
that only one of the particles composing the system is
directly struck by the external probe. This produces a relatively
small perturbation of the system which however responds as a whole
because the knocked particle is linked to the other ones. The total
response is related to the characteristics of the interaction among
the particles. 

In the electron excitation of the atomic nuclei, the region where the
nuclear response is quasi-elastic is well identified. The inelastic
electron scattering cross section shows an evident peak at higher
energies than those of the giant resonance region. The position
of this peak is located at values very close to $|{\bf q}|^2/2M$,
where ${\bf q}$ is the momentum transferred from the electron
to the nucleus and $M$ is the nucleon mass. The width of the peak
increases when the mass number of the target nucleus increases. These
two facts lead to the interpretation of the peak as produced by the
elastic scattering of the electron with a single nucleon. An ideal
experiment on a free nucleon produces a narrow  peak 
positioned  exactly at $|{\bf q}|^2/2M$. Understanding the
deviation from this situation means understanding the nuclear electron
excitation mechanism and the nuclear many-body dynamics.
%
%
\begin{figure}[ht]
\vspace{-2cm}
\begin{center}                                                                
\leavevmode
\epsfysize = 400pt
\hspace*{.45cm}
\makebox[0cm]{\epsfbox{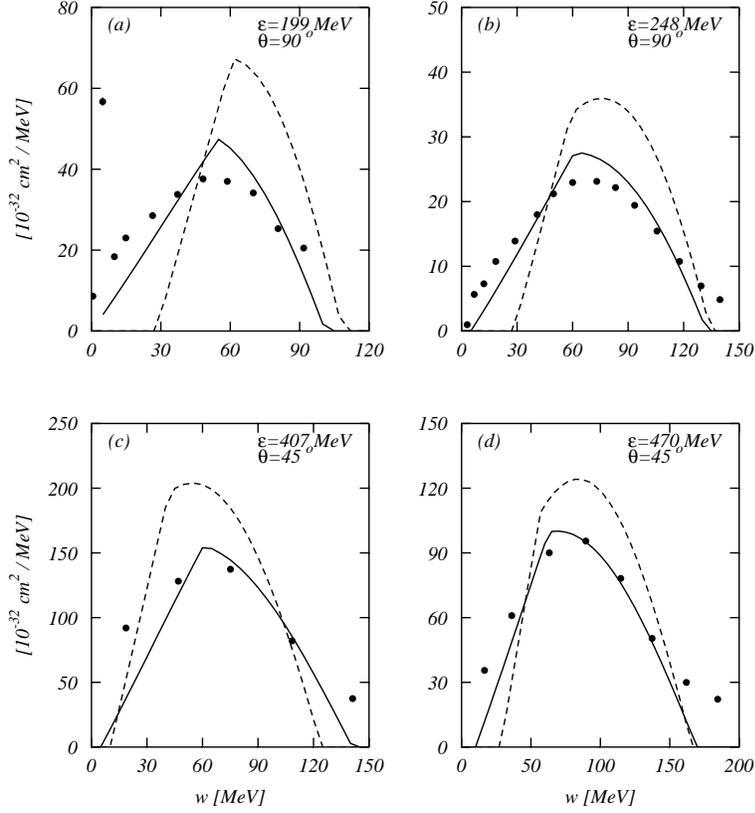}}
\end{center}
\vspace{-1.5cm}
\caption[]{\small 
FG model results compared with the experimental data of
refs. \cite{Yat93,Wil97} for the $^{40}$Ca nucleus. The electron
energies $\varepsilon$ and the scattering angles $\theta$ are
indicated in the figures. The dashed lines show the results obtained
by using the values of $k_{\rm F}$=251~MeV/$c$ and $\langle E
\rangle$=28~MeV fixed in ref. \cite{Mon71}. The full lines have been
calculated changing the values of the Fermi momentum and of the average
binding energy to reproduce at best the experimental data: 
(a) $k_{\rm F}$=340~MeV/$c$ and $\langle E \rangle$=0~MeV, 
(b) $k_{\rm F}$=330~MeV/$c$ and $\langle E \rangle$=5~MeV, 
(c) $k_{\rm F}$=330~MeV/$c$ and $\langle E \rangle$=5~MeV and
(d) $k_{\rm F}$=320~MeV/$c$ and $\langle E \rangle$=90~MeV.
}
\label{moniz}
\end{figure}

The first electron scattering experiment aimed to probe the nuclear QE
response was done at the Stanford Mark III linear accelerator in the
early seventies \cite{Mon71}. The data were analyzed within a Fermi
gas (FG) model \cite{Mon69}, i.e. considering the nucleus as an
infinite system of non interaction fermions, and the corresponding
fits were obtained by changing the two parameters of the model: the
average nucleon interaction energy, $\langle E \rangle$, and the Fermi
momentum, $k_{\rm F}$. The position of the peak is fixed by $\langle E
\rangle$, while both height and width are determined by the Fermi
momentum. The interesting result was that the three quantities could
be reproduced so well by two parameters. 

The aim of this experiment
work was to obtain an empirical estimate of the value
of $k_{\rm F}$. 
The commonly accepted value of $k_{\rm F}$ (about
270~MeV/$c$) is obtained from the empirical value of nuclear matter
density, $\rho=0.17$~fm$^{-3}$, fixed by considering the nucleus a
homogeneous sphere of radius $R=r_0 A^{\frac{1}{3}}$, with
$r_0=1.18$~fm given by elastic electron scattering experiments. The
fit procedure used to reproduce the QE data provided for heavy nuclei
a roughly constant value of $k_{\rm F}$ of about 260~MeV/$c$, in good
agreement with the value obtained with the usual procedure.

The success of this description of the data within a FG model suppressed
interest in the investigation of the QE peak for about ten years,
until the data of the separated charge and current responses became
available \cite{Alt80}. The FG model which
successfully explained the cross
section was unable to describe separately its two components
\cite{Alt80}. 

Looking back at the analysis of the  pioneering Stanford experiment  we
find that what is nowadays considered interesting physics is hidden in
the two parameters of the fit. The nuclear structure problem and the
interplay between single-particle and many-body dynamics have been
obscured by the fit procedure. 

In fig. \ref{moniz} we compare the FG results with the recent $^{40}$Ca
data taken at Bates \cite{Yat93,Wil97}. The values of $\langle E
\rangle$ and $k_{\rm F}$ defined in ref. \cite{Mon71} produce cross
sections well above the data (see dashed curves). On the other hand,
if $\langle E \rangle$ and $k_{\rm F}$ are fixed to reproduce one data
set (full lines) the other ones are not well described.

The FG model provides a qualitative description of the phenomenon, but
to obtain a quantitative description of the different sets of data it
is necessary to go beyond this model and treat explicitly those
degrees of freedom that are hidden in the effective values of the two
parameters of the model. 

The aim of this article is to discuss the successes and failures of the MF
description of the electron induced QE excitation of
medium and heavy nuclei. The discussion will be focused on what
can be learnt about the interaction between the electromagnetic
probe and the nucleus and the dynamics of the nuclear excitation.
In the presentation we shall try to separate the problems
concerning the reaction mechanism from those regarding the nuclear
structure. This separation is somewhat artificial but it is useful to
better determine and clarify the various approximations and uncertainties
done in the study of the QE peak.

The description of the reaction mechanism is based upon the Plane Wave
Born Approximation and the use of the traditional
electromagnetic one-body operators, and we shall discuss the limits of
all these approximations considering the distortion of the electron
wave functions, the presence of Meson Exchange Currents and the
relativistic corrections to the one-body operators.

The nuclear structure problem is tackled within the non-relativistic
formalism. The basic model we shall adopt is the Continuum Shell Model,
where the nucleons move independently from each other in a
finite size average potential. We shall present, however, results
obtained by describing the nucleus as an infinite system of non
interacting nucleons. In addition to these MF descriptions of the
excitation we shall also discuss results obtained with the Random
Phase Approximation theory.

\section{The electron-nucleus interaction}
\label{em}

Our understanding of the electromagnetic processes is certainly deeper
than that of those processes dominated by the strong
interaction. There are however various problems not very well solved,
or defined, in the description of the interaction between electrons
and nuclei. In this section we shall discuss some of them. We shall
present first what we call the standard derivation of the inelastic
electron nucleus cross section. Then we shall discuss the validity of
the approximations done in this approach, with special attention to
their influence on the study of the QE excitation. In particular, the
role of the nucleon form factors, the relativistic corrections and the
Meson Exchange Currents are discussed.

\subsection{The standard treatment}
\label{st}
In this section we  present the basic points of the usual derivation
of the electron scattering cross section off atomic nuclei. More
detailed presentations of this derivation can be found in various
review articles (see, e.g., \cite{deF66,Don75,Cio80,Dre89,Bof96}). 

Throughout the paper we shall use the Bjorken and Drell \cite{Bjo64}
metric and the convention that the repeated indexes of a four-vector
imply a sum on them: $a_\mu b^\mu=a_0 b^0 - {\bf a \cdot b}$.

Using the standard Feynman rules, the scattering amplitude
for a single photon exchange process can be written as \cite{Bjo64}:
\begin{equation}
\label{amplitude1}
\langle f,k_f | S | i,k_i \rangle \, = \,
-i\int {\rm d}^4x \, [ e {\overline \psi}_f (x) \gamma_\mu
A^\mu(x) \psi_i (x) ] \, .
\end{equation}
In this equation the initial (final) state of the system is
defined as the tensor product between the initial (final) state of the
electron and of the nucleus. As we work in the Plane Wave
Born Approximation (PWBA), the
electron wave functions are plane wave solution of the free Dirac equation:
\begin{equation}
\label{elwave}
\psi (x) \, =\,  \sqrt{ \frac{m}{\varepsilon V}} \,
u(k,s)\, e^{-ik^\mu x_\mu} \, .
\end{equation}
where $u$ is one of the positive energy
four-component Dirac spinors \cite{Bjo64}. The
electron states are characterized by the values of the four-momentum
$k \equiv \{\varepsilon,{\bf k}\}$. The final formulae will be
independent of the normalization volume $V$. 

In eq. (\ref{amplitude1}) we have indicated with $A^\mu$ the classical
electromagnetic field generated by the nucleus. This should satisfy
the field equation: 
\begin{equation}
\label{field}
\Box A^\mu(x) \, = \, e \langle f | J^\mu(x) | i \rangle \, ,
\end{equation}
where $\Box \equiv \nabla^2- \partial^2 / \partial t^2$ and $J^\mu$ is
the nuclear current operator. The calculation now proceed by inserting
the solution of eq. (\ref{field}) and the expression of the electron
wave functions (\ref{elwave}) in eq. (\ref{amplitude1}) which can be
rewritten as:
\begin{equation}
\langle f,k_f | S | i,k_i \rangle \, = \, 
-ie^2 \langle k_f | j_\mu | k_i \rangle 
\frac {g^{\mu\nu}}{q^\beta q_\beta}
\int {\rm d}^4x \, e^{-iq_\alpha x^\alpha} 
\langle f | J_\nu(x) | i \rangle \, ,
\end{equation}
where the four-momentum transfer is defined as 
$q^\nu=k^\nu_i-k^\nu_f$ and the electron current is given by:
\begin{equation}
\langle k_f | j_\mu | k_i \rangle \, = \, 
\frac{m}{V} \sqrt{\frac{1}{\varepsilon_i \varepsilon_f}} \,
{\overline u}(k_f,s_f) \, \gamma_\mu  \, u(k_i,s_i) \, .
\end{equation}
The nuclear current is supposed to be described by an operator local
in ${\bf r}$. Its time dependence is given by: 
\begin{equation}
\label{heisenberg}
J_\mu(t,{\bf r}) \, = \, e^{iHt} \, J_\mu({\bf r}) \, e^{-iHt} \, .
\end{equation}
After performing the integration on the time variable the scattering
amplitude can be expressed as:
\begin{equation}
\label{amplitude2}
\langle f,k_f | S | i,k_i \rangle \, = \, 
-ie^2 \, 2\pi \, \delta(E_i - E_f - \omega) \,
\langle k_f | j_\mu | k_i \rangle \, \frac {1}{q^\nu q_\nu} \,
\langle f | J^\mu({\bf q}) | i \rangle \, ,
\end{equation}
where the $E_i$ and $E_f$ indicate, respectively,
the initial and final energy of the nucleus
and $\omega$ is the energy transferred by the electron. The expression
(\ref{amplitude2}) depends explicitly on the Fourier transform of the
nuclear current
\begin{equation}
\label{currentf}
J_\mu({\bf q}) \, = \, \int {\rm d}^3x \, 
e^{i{\bf q \cdot r}} \, J_\mu({\bf r}) \, .
\end{equation}
The nuclear current should satisfy the continuity equation:
\begin{equation}
\label{cont1}
\partial_\mu J^\mu(x)=0 \, ,
\end{equation}
which, using the explicit time dependent expression (\ref{heisenberg}),
can be written, in coordinate space, as :
\begin{equation}
\label{cont2}
\nabla \cdot {\bf J}({\bf r}) \, = \, - i \omega J^0 ({\bf r})
\end{equation}
while in momentum space it assumes the form:
\begin{equation}
\label{cont3}
{\bf q} \cdot  {\bf J}({\bf q}) \, = \, \omega \, \rho ({\bf q}) \, ,
\end{equation}
where we have used $ J^0 \equiv \rho$. The continuity equation imposes
a relation between the four components of the nuclear current,
therefore only three of them are independent. It is necessary to
choose the independent components to be used to express the scattering
amplitude. It is convenient to work in a reference system where the
quantization $z$-axis points along the direction of the momentum
transfer ${\bf q}$. Eq. (\ref{cont3}) shows that the component of the
current along this axis, the longitudinal component $J_z$, can be
expressed in terms of the charge operator $\rho$. Therefore the
scattering amplitude can be expressed in terms of $\rho$ and of two
components orthogonals to the direction of ${\bf q}$. These are called
transverse components and they are usually expressed in spherical
coordinates: 
\begin{equation}
{\bf J}_{\pm}=\mp \frac{1}{\sqrt{2}} 
\left({\bf J}_x \pm i {\bf J}_y  \right) \, .
\end{equation}

To calculate the cross section one has to square the scattering
amplitude, to average on the initial states and to sum over the final
ones. Since the electron energies $\varepsilon$ are much bigger than
its rest mass $m$, it is common to use the so called ultrarelativistic
approximation consisting in neglecting the terms depending on $m$. The
calculation is rather long but it can be done without any further
approximation. The cross section assumes the expression:
\begin{equation}
\label{xsect}
\frac {{\rm d}^2\sigma}{{\rm d}\Omega {\rm d}\varepsilon_f} \, = \,
\sigma_{\rm Mott} \,
\left\{ \left(\frac{q^\mu q_\mu}{q^2}\right)^2 R_L(q,\omega)
+ \left(\tan^2 \frac{\theta}{2}-\frac{q^\mu q_\mu}{2q^2}  
\right)R_T(q,\omega)
\right\} \, ,
\end{equation}
where $\theta$ is the scattering angle, defined as the
angle between the directions of ${\bf q}$ and ${\bf k}_f$,  
$q=|{\bf q}|$ and $\sigma_{\rm Mott}$ is the Mott cross section
\begin{equation}
\label{mott}
\sigma_{\rm Mott} \, = \,
\left(\frac {\alpha \cos(\theta/2)}{2\varepsilon_i \sin^2(\theta/2)}   
\right)^2 \, ,
\end{equation}
where we have indicated with $\alpha$ the fine structure constant.

In the expression (\ref{xsect})  the cross section is composed of two parts,
each of them formed by a kinematical term and a response function containing
the full nuclear structure information.
The longitudinal and transverse response functions are respectively
defined as:
\begin{eqnarray}
\label{respl}
R_L(q,\omega)&=& \overline {\sum_i}  \sum_f
\delta(E_f-E_i-\omega) |\langle f | \rho(q) | i \rangle|^2  \\
\label{respt}
R_T(q,\omega)&=& \overline {\sum_i} \sum_f
\delta(E_f-E_i-\omega) 
\left( |\langle f | J_+(q) | i \rangle|^2 \, + \, 
|\langle f | J_-(q) | i \rangle|^2 \right) \, .
\end{eqnarray}
The bar on the sum on $i$ indicates the average on the initial states.

If the expression (\ref{xsect}) of the cross section is valid it is
possible to obtain an experimental separation of the longitudinal and
transverse responses. To achieve this goal it is necessary to perform
a set of  measurements at fixed $\omega$ and $q$ but for different
values of the scattering angle $\theta$. The ratio of the cross
section and $\sigma_{\rm Mott}$ should be plotted against
$v_T=\tan^2(\theta/2)-(\omega^2-q^2)/(2q^2)$ and all the points should
lie on a straight line. The slope of this line gives $R_T$ while the
intersection with the $y$ axis provides $R_L$. This procedure is
called the Rosenbluth separation \cite{Ros50}.

The nuclear responses are calculated under the hypothesis that the
nucleus makes a transition  from one state of definite angular
momentum $|i\rangle \equiv |J_i M_i\rangle$ to another one $| f
\rangle \equiv |J_f M_f\rangle$. For this reason it is convenient to
rewrite eq. (\ref{currentf}) by making a multipole expansion of the
exponential. Considering the parity of the nuclear final state, it is
possible to separate the electric and magnetic transitions and at the
end one has to deal with three type of multipole operators: the
Coulomb operators producing transitions of natural parity, 
\begin{equation}
\label{mlm}
M_{JM}(q) \, = \, \int {\rm d}^3r \, j_J(qr)
\, Y_{JM}(\hat{r})\, \rho({\bf r}) \, ,
\end{equation}
the electric current operators which give rise also to transitions of
natural parity,
\begin{equation}
\label{telm}
T^{\rm E}_{JM}(q) \, = \, \frac{1}{q} \int {\rm d}^3r \,
\left\{ \nabla \times \left[j_J(qr){\bf Y}^M_{JJ}(\hat{r}) \right]
\right\} \cdot {\bf J}({\bf r}) \, ,
\end{equation}
and the magnetic current operators producing transitions of
unnatural parity,
\begin{equation}
\label{tmlm}
T^{\rm M}_{JM}(q) \, = \, \int {\rm d}^3r \,
j_J(qr) \, {\bf Y}^M_{JJ}(\Omega) \cdot {\bf J}({\bf r}) \, .
\end{equation}
In the eqs. (\ref{mlm})-(\ref{tmlm}) $j_J$ indicates a spherical
Bessel function, $Y_{JM}$ is a spherical harmonics and ${\bf
Y}^M_{JL}$ is a vector spherical harmonics \cite{Edm57}.

We shall deal with scattering from doubly closed shell nuclei,
therefore $J_i=0$ and then the value of the angular momentum of the
final state correspond to the multipolarity of the transition. The
nuclear responses can be written as:
\begin{eqnarray}
\label{rl1}
R_L(q,\omega) &=&  4 \pi \sum_\beta \sum^\infty_{J=0} \,
|\langle \beta,J \parallel M_J(q) \parallel 0 \rangle|^2 \,
\delta(E_\beta-\omega) \, , \\
\label{rt1}
R_T(q,\omega) &=& 4 \pi \sum_\beta \sum^\infty_{J=1} \,
\left(
|\langle \beta,J \parallel T^{\rm E}_J(q) \parallel 0 \rangle|^2 +
\right. \nonumber  \\
&& \left. \hspace{2cm}
|\langle \beta,J \parallel T^{\rm M}_J(q) \parallel 0 \rangle|^2 
\right) \, \delta(E_\beta-\omega) \, ,
\end{eqnarray}
where $\beta$ indicates all the quantum numbers characterizing the
nuclear final state other than $J$. 

The only elements which remain to be specified are the explicit
expressions of the nuclear electromagnetic operators $\rho$ and ${\bf
J}$. Up to now the only hypothesis made on these operators has been
done in eq. (\ref{heisenberg}). Since in general the nuclear many-body
wave functions are described using nucleonic degrees of freedom in a
non relativistic framework, the standard treatment considers the
electromagnetic operators produced by non relativistic pointlike
nucleons. This means that the nuclear charge is composed by the charge
of the nucleons and the current is composed by the movement of the
charged nucleons inside the nucleus (convection current) and by the
sum of the magnetic currents associated with the nucleons spin
(magnetization current). The expressions of these operators are:
\begin{equation}
\label{charge1}
\rho({\bf r}) \, = \, 
\sum^A_{k=1} \frac{1+\tau^3_k}{2} \, 
\delta({\bf r}-{\bf r}_k) \, ,
\end{equation}
for the charge operator,
\begin{equation}
\label{convection1}
{\bf j}^{\rm C}({\bf r}) \, = \, 
\sum^A_{k=1} \frac{-i}{2M_k} \, \frac{1+\tau^3_k}{2} \,
\left[ \delta({\bf r}-{\bf r}_k)\nabla_k
+ \nabla_k\delta({\bf r}-{\bf r}_k) \right] \, ,
\end{equation}
for the convection current operator, and
\begin{equation}
\label{magnetization1}
{\bf j}^{\rm M}({\bf r}) \, = \, 
\sum^A_{k=1} \frac{1}{2M_k} \,
\left(\mu^{\rm P}\frac{1+\tau^3_k}{2} + 
\mu^{\rm N}\frac{1-\tau^3_k}{2} \right)
\nabla \times \delta({\bf r}-{\bf r}_k) \, \nsigma_k \, ,
\end{equation}
for the magnetization current operator. In the previous equations
$M_k$ indicates the rest mass of $k$-th nucleon, $\mu^{\rm P}$ and
$\mu^{\rm N}$ are the anomalous magnetic moment of the proton and the
neutron respectively, $\nsigma_k$ is the Pauli spin matrix of the 
$k$-th nucleon and $\tau^3_k=1$ or -1 according the $k$-th nucleon
to be a proton or a neutron. 

The expressions (\ref{charge1}), (\ref{convection1}) and 
(\ref{magnetization1}) of the one-body nuclear charge and current are 
obtained as Fourier transform of a specific non relativistic reduction
of a nucleon current of the type:
\begin{equation}
\label{relc1}
J^{\mu}(P_f S_f ,P_i S_i) \, = \,
\overline{u}(P_f,S_f) \, \gamma^{\mu} \, u(P_i,S_i) \, ,
\end{equation}
where the $u$ are the nucleon Dirac spinors and $P$ and $S$ are
respectively the four momentum and the spin of the nucleon. The
procedure leading to the expressions
(\ref{charge1})-(\ref{magnetization1}) consists in retaining the zero
and first order terms of an expansion in powers of
$q^\nu=P^\nu_f - P^\nu_i$.

\subsection{Coulomb distortion}
\label{coulomb}
The possibility of making an experimental separation of the
longitudinal and transverse response functions is very appealing
because it allows one to investigate separately the electron nucleus
interaction and the nuclear excitation mechanism. The same nuclear
model should be used to describe both responses, the only
difference between them is due to the electromagnetic operator.
%
%
\begin{figure}
\begin{center}                                                                
\leavevmode
\epsfysize = 500pt
\hspace*{-.6cm}
\makebox[0cm]{\epsfbox{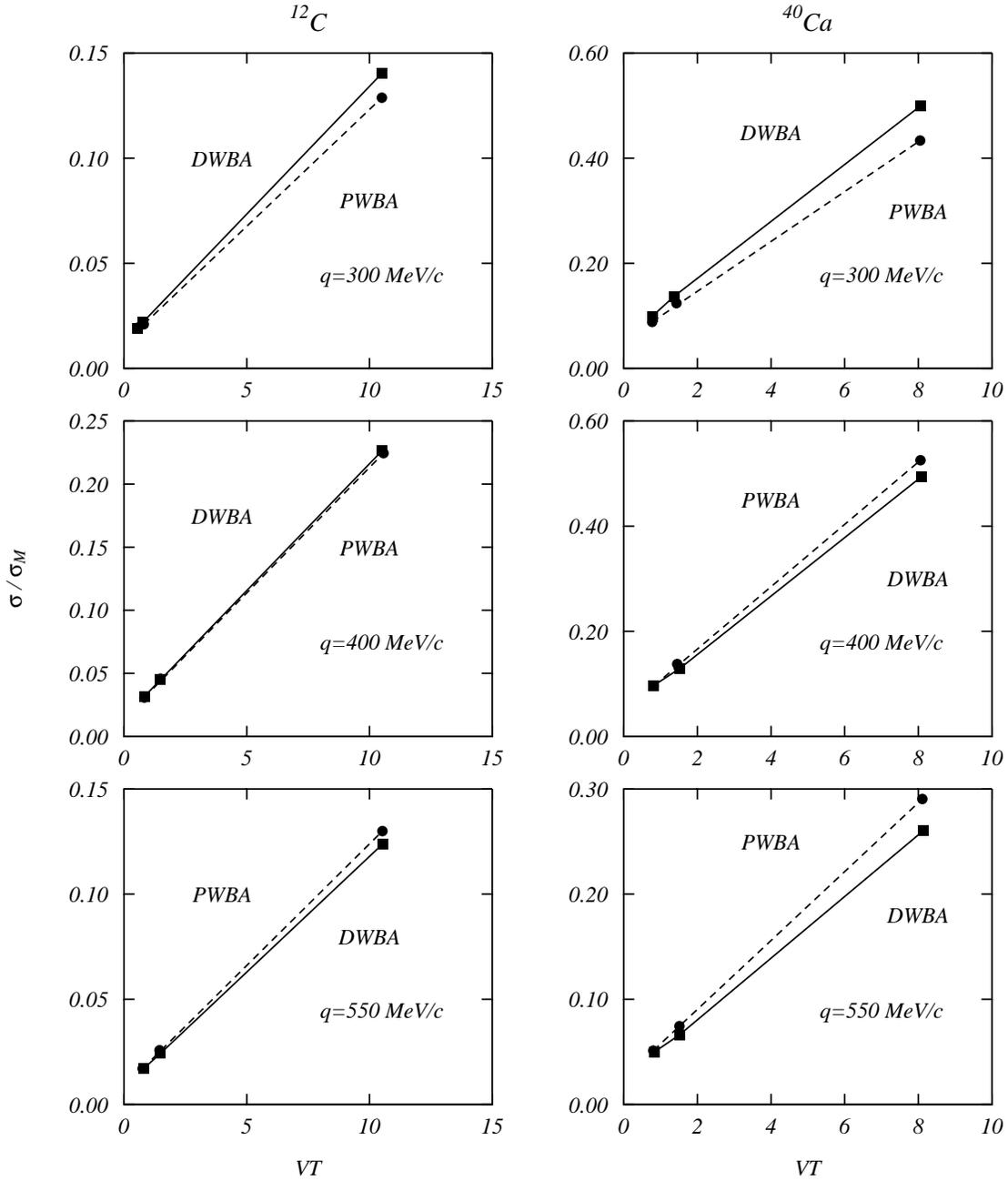}}
\end{center}
\vspace{-1cm}
\caption{\small Rosenbluth plot for $^{12}$C(e,e')$^{12}$C
and $^{40}$Ca(e,e')$^{40}$Ca reactions at the fixed excitation energy 
$\omega=300$~MeV. We have indicated with $\sigma_M$ the Mott cross section.
Note that the DWBA points are not perfectly aligned.
}
\label{dwba}
\end{figure}

We have seen that the Rosenbluth separation is possible if the
expression (\ref{xsect}) of the cross section is valid. In this
expression the kinematics is well separated from the nuclear structure
since the PWBA has been used, in other words, since it has been
assumed that the electron wave functions are plane wave solutions of
the Dirac equation. In this approach the effects of the
electromagnetic field of the nucleus on the motion of the electron are
neglected. The validity of this approximation is quite good for few
body systems or light nuclei, but it starts to become questionable
already for nuclei as heavy as $^{12}$C \cite{deF66,Cio80,Hei83}.

It is possible to solve the Dirac equation for an electron moving in a
static Coulomb potential \cite{Yen54}. This provides the exact
solution for the elastic scattering cross section. The situation is
more difficult to handle in the case of inelastic scattering since the
electromagnetic field generated by the nucleus is not any more static
because it is produced by a transition from the ground to the excited
state. In the most sophisticated treatment developed up to now, the
one-photon exchange scattering amplitude (\ref{amplitude1}) is
calculated using the electron wave functions obtained by solving the Dirac
equation with a Coulomb potential generated by the target nucleus
ground state. This treatment is called
Distorted Wave Born Approximation (DWBA) and its validity in the
description of low lying states has been widely studied \cite{Hei83}. 

There are two major effects that the DWBA adds to the PWBA
description. The first is produced by the attraction between the
positively charged nucleus and the negatively charged electron. The
asymptotically  measured initial and final momenta $k_i$ and $k_f$ are
smaller than the effective momenta of the electron around the
nucleus. This is an interpretation of the phenomenon based upon a PWBA
description of the process, since DWBA wave functions are not
eigenstates of the momentum operator of the electron. 
In this simplified picture
the incoming electron is accelerated by the nucleus while the
scattered electron is decelerated. This produces a transferred
momentum larger than that of the  PWBA. An empirical expression of the
effective momentum transfer is obtained in the eikonal approximation
\cite{Hei83}:
\begin{equation}
\label{effmom}
q_{\rm eff} \, = \, q 
\left(1+1.5 \frac{Z \alpha \hbar c} {\varepsilon_i R_{\rm eq}} 
\right)
\end{equation}
with $R_{\rm eq}=1.24\, A^{1/3}$. The difference between the PWBA
momentum transfer and the effective one increases with the
charge $Z$ of the target nucleus and diminishes when the electron 
energy $\varepsilon_i$ increases.

The second DWBA effect is of quantum mechanical nature and consists in
the focusing of the electron wave function onto the nucleus causing
an increase of the flux and an enhancement of the cross section. This
effect is partially compensated by the reduction of the cross section due
to the $q_{\rm eff}$ shift.

In the DWBA expression of the cross section it is not possible to
separate the nuclear from the electron variables. Furthermore there
are interference terms between longitudinal and transverse
amplitudes. In a Rosenbluth plot the DWBA cross sections do not lie
on a straight line. If the distortion effects are small, however, it
is possible to correct the experimental cross sections in order to have 
PWBA equivalent cross sections and to use them to make the Rosenbluth 
separation.

Distortion effects on the QE excitation have been studied for $^{12}$C
and $^{40}$Ca nuclei in ref. \cite{Co'87a}, where a full DWBA
calculation has been done, and for $^{40}$Ca and $^{208}$Pb nuclei
in ref. \cite{Tra88,Tra93}, where an analytical expansion of the
amplitudes up to $(Z\alpha)^2$ has been used.

A first finding of these studies is that the DWBA cross sections 
without corrections are
rather well aligned on a Rosenbluth plot. Within the experimental 
uncertainty their alignment is compatible with an exact
straight line (see fig. \ref{dwba}). For this reason the alignment of
the cross sections on a Rosenbluth plot cannot be considered a proof
of the validity of PWBA.

The major effect of the distortion on a Rosenbluth plot consists in
a rotation of the line around the intersection point with the $y$
axis. This does not modify very much the value of the longitudinal
response but it has a rather big effect on the value of the transverse
response function. These considerations are valid for a fixed value of
the excitation energy.

In ref. \cite{Tra88} the full energy dependence of the responses has
been studied. The distortion effects on $^{40}$Ca consists mainly of
a shift to higher energies and it can be rather well described in terms
of $q_{\rm eff}$. The situation in $^{208}$Pb is more complicated since
the focusing of the wave function plays an important role.

The effects of the distortion are  small enough to be considered a
correction to the PWBA, therefore it is reasonable to extract the
responses from the cross section using the procedure outlined above.

\subsection{Nucleon electromagnetic form factors}
\label{formfact}
%
%
\begin{figure}
\begin{center}                                                                
\leavevmode
\epsfysize = 500pt
\hspace*{-0.7cm}
\makebox[0cm]{\epsfbox{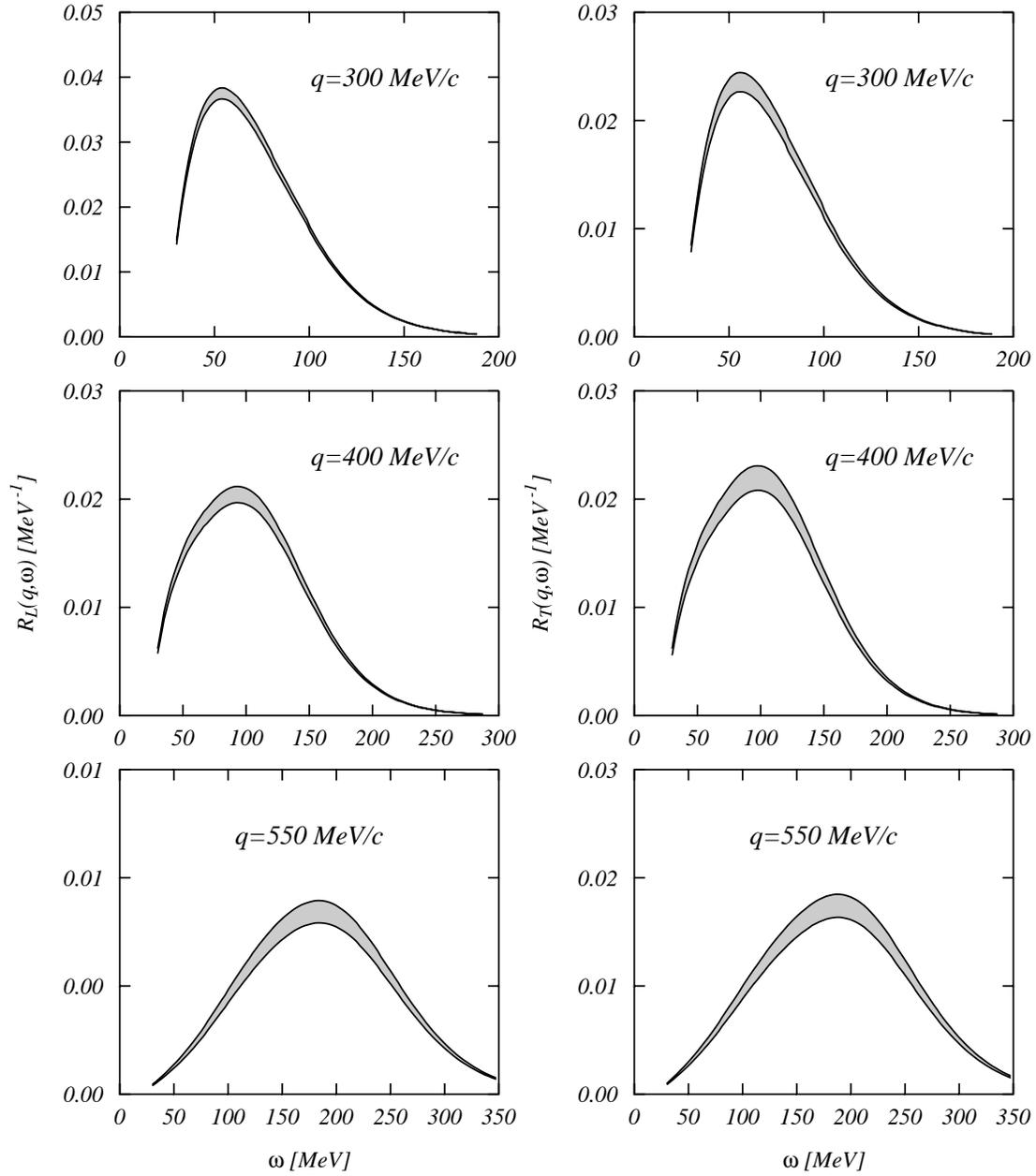}}
\end{center}
\vspace{-1.cm}
\caption{\small $^{12}$C response functions. The bands show the uncertainty 
related to the choice of the  the nucleon electromagnetic
form factors.
}
\label{form}
\end{figure}
In the standard derivation, the nuclear current has been obtained
under the assumption of pointlike nucleons. In the QE excitation the
values of the energies and momenta involved are such that the nucleon
internal structure plays a noticeable role. This fact is taken into
account by inserting nucleon electromagnetic form factors in the
expression of the current. For a nucleon of type $\tau$ (either proton
or neutron) with mass $M$, eq. (\ref{relc1}) becomes:
\begin{equation}
J^{\mu}_\tau(P_f S_f,P_iS_i) \, = \,
\overline{u}(P_f,S_f)
\left[ F^\tau_1(q^2)\gamma^{\mu}
+ i\kappa^\tau \frac{F^\tau_2(q^2)}{2M}\sigma^{\mu\nu}q_{\nu}\right]
u(P_i,S_i)
\end{equation}
where
\begin{eqnarray}
\sigma^{\mu \nu}= 
\frac{i}{2}(\gamma^\mu \gamma^\nu - \gamma^\nu\gamma^\mu)
\end{eqnarray}
and $\kappa^\tau$ is set equal to the anomalous part of the magnetic
moment in units of nucleon magnetons (1.79 for proton and -1.91 for
neutron). The terms $F^\tau_1$ and $F^\tau_2$ are the Dirac form
factors and they are normalized such as:
\begin{eqnarray}
F^p_1(0)=1 \,\,\,\,&&\,\,\,\, F^p_2(0)=1 \\
F^n_1(0)=0 \,\,\,\,&&\,\,\,\, F^n_2(0)=1 \, .
\end{eqnarray}
For a more direct comparison with the experiment it is convenient 
to use the Sachs form factors related to the Dirac ones 
by the expressions:
\begin{eqnarray}
G^\tau_{\rm E} &=& 
F^\tau_1+\kappa^\tau \frac{q_\nu q^\nu}{4M^2}F^\tau_2 \\
G^\tau_{\rm M} &=& F^\tau_1+\kappa^\tau F^\tau_2 \, . 
\end{eqnarray}

The effect of the nucleon finite size is implemented by means of a
convolution of the pointlike currents
(\ref{charge1})-(\ref{magnetization1}) with these nucleon form
factors. Then, it is more convenient to work in $q$ space where the
nucleon form factors are simply multiplicative factors. 

The problem is now to find a functional dependence of the form factor
in terms of $q$ and $\omega$ which is able to reproduce the elastic
electron-nucleon data. Unfortunately the data are not selective enough
to define a unique parameterization of this functional
dependence. Different type of form factors can reproduce the elastic
scattering data with the same degree of accuracy.

The sensitivity of the nuclear responses to the different choices of
the electromagnetic nucleon form factors has been studied in
ref. \cite{Ama93b}. In that article we performed calculations of the
QE responses with the same nuclear structure input but considering the
nucleon form factors of refs. \cite{Jan66,Ber72,Iac73,Hoe76,Sim80}. We
found that the momentum and energy domain of the QE peak lies in a
region of strong variations of the nucleon electromagnetic form
factors, and the changes of the nuclear responses can be as big as
12\% at the peak values (see fig. \ref{form}). 

Some of the parametrizations we have used are  old and have been ruled
out by more modern measurements. Probably the 12\% uncertainty we have
quoted slightly overestimates the actual uncertainty on the form
factor. In any case the study quoted above
points out the need of specifying the
nucleon form factor used when one compares the results of QE
calculations with the experimental data. The ambiguities in the choice
of the nucleon form factors set an intrinsic uncertainty threshold of
the theoretical calculations.

\subsection{Relativistic corrections}
\label{relem}
We have already said that the expressions of the one-body operators
(\ref{charge1})-(\ref{magnetization1}) are obtained by expanding the
relativistic operator (\ref{relc1}) in powers of $q$, $\omega$ and
$P$, and retaining the zero and first order terms.

\begin{figure}
\begin{center}                                                                
\leavevmode
\epsfysize = 500pt
\hspace*{-0.85cm}
\makebox[0cm]{\epsfbox{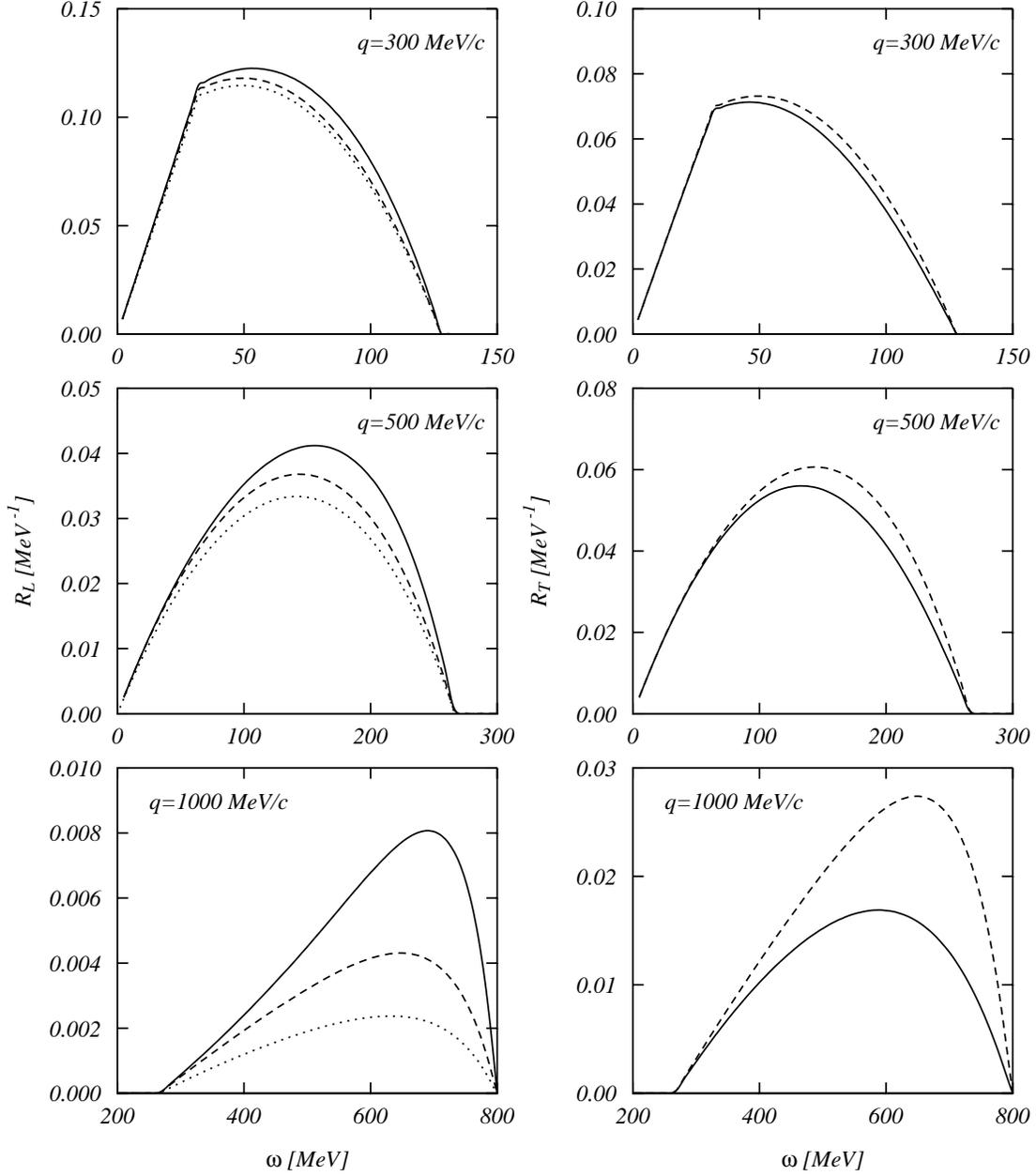}}
\end{center}
\vspace{-1cm}
\caption{\small The lines in this figure have all been obtained using
the non relativistic FG model. The dashed lines present the result
obtained by using the usual non relativistic one-body operators. The
dotted curves include the Darwin-Foldy term (only in the longitudinal
response). The full lines have been obtained considering the
relativistic corrections to the operators.}
\label{relop}
\end{figure}

In ref. \cite{Voy62} it has been pointed out that a consistent treatment
of both the charge and the current operators, up to the same order in 
the expansion, produces a correction factor for the charge operator. 
This factor is called the Darwin-Foldy term
\begin{equation}
\label{darwin}
\displaystyle
f_{\rm DF} \, = \, \left( 1+\frac{q^2-\omega^2}{4M^2} \right)^{-1/2}
\end{equation}
and it multiplies the electric Sachs form factor in the calculation of
$R_L$. This correction reduces the longitudinal response as is
shown in fig. \ref{relop}.
%
%

A new non relativistic expansion of the on-shell electromagnetic
one-body current has been proposed in ref. \cite{Ama96a} for use in
the region of the QE peak at high $q$ values. The idea of this
approach is to obtain expressions of the current operators which are
not truncated in powers of $q$ and $\omega$ and hence applicable at
high values of energy and momentum transfer. 

In a FG model the value of the momentum of a bound nucleon is at most
of the order of the Fermi momentum $P\sim k_{\rm F}$. Therefore for a
typical value of $k_{\rm F}=250$~MeV/$c$ we have $P/M \le 1/4$ and
hence an expansion of the current in powers of $P/M$ is justified.

The details of the calculation can be found in
refs. \cite{Ama96a,Ama96b,Jes98} and the final expressions of the
charge and transverse current operators  up to first order in $P/M$ are:
\begin{eqnarray}
\label{relcharge}
\rho ({\bf q},\omega) &=& \frac{q}{\sqrt{q^2-\omega^2}}\, G_{\rm E}
        +i \, \frac{ \left(G_{\rm M}- \frac{1}{2} G_{\rm E} \right)}
             {M\sqrt{4M^2+q^2-\omega^2}}
        ({\bf q}\times {\bf P})
        \cdot\nsigma \, , \\
\label{relcurrent}
{\bf J}^T ({\bf q},\omega) 
     &=& \frac{\sqrt{q^2-\omega^2}}{q}
         \left\{ i\frac{G_{\rm M}}{2M}(\nsigma\times {\bf q})+ 
          \frac{G_{\rm E}}{M} \left({\bf P}
                 -\frac{{\bf q} \cdot {\bf P}}{q^2}
               {\bf q} \right) 
         \right\} \, .
\end{eqnarray}
The first piece of the charge operator is the usual non relativistic
contribution $G_{\rm E}$ multiplied by a factor larger than one. 
This factor produces (see fig. \ref{relop}) an increase of the non
relativistic result which goes in the opposite direction with respect
to the traditional Darwin-Foldy correction, and it can be directly
implemented as a multiplicative factor in the non-relativistic models
of the reaction. The second piece is the charge of the spin-orbit
operator, which is slightly different from the one arising in the usual
expansions. Finally, the transverse current operator contains the two
traditional pieces, the magnetization and convection current
operators, but both are corrected by a multiplicative factor smaller
than one.

The results presented in fig. \ref{relop} show big relativistic
effects at momentum transfer larger than 500~MeV/$c$. In these
situations the value of the momentum of the emitted nucleon is
comparable with its rest mass $M$, therefore a relativistic treatment
of the motion of the emitted nucleon is necessary. This involves the
nuclear structure part of the problem and we shall discuss this point
more in detail in section \ref{relstr}.

\subsection{Two-body currents}
\label{mec}
We have treated the problem of the electron excitation considering
the nuclear current produced only by individual nucleons. 
This limitation is
inconsistent already at the level of the standard derivation. Inserting
the expression (\ref{heisenberg}) into the continuity equation
(\ref{cont1}) one obtains:
\begin{equation}
\label{cont4}
\nabla \cdot {\bf J}({\bf r})\, = \, - i \left[H,\rho \right] \, = \,
- i \left[T+V,\rho \right] \, ,
\end{equation}
where $T$ is the kinetic energy and $V$ the interaction term.

It can be shown that by using the expressions (\ref{charge1}) and 
(\ref{convection1}) for the charge and convection current operators
the equation
\begin{equation}
\nabla \cdot {\bf j}^{\rm C}({\bf r})\, = \, - i \left[T,\rho \right]
\end{equation}
is satisfied. It is obvious that a new term must be added to the
one-body current in order to verify the continuity equation with the
full hamiltonian. This new term, ${\bf J}^{\rm TB}$, must satisfy
the equation
\begin{equation}
\nabla \cdot {\bf J}^{\rm TB}({\bf r})\, = 
\, - i \left[V,\rho \right]
\end{equation}
and it is a two-body operator.

The continuity equation shows the theoretical needs for including
two-body currents, but it does not define them in an unique
way, because all the operators which are divergenceless, for example 
the one-body magnetization current (\ref{magnetization1}), are not
restricted by this equation. 

Since there is not a unique way to derive these two-body current
operators, different approaches have been proposed in the
literature. In any case, it is commonly accepted that the most
important terms are those related to the exchange of mesons
and more specifically to the exchange of a single pion. 

\begin{figure}
\vspace{-9cm}
\begin{center}                                                                
\leavevmode
\epsfysize = 750pt
\hspace*{-.8cm}
\makebox[0cm]{\epsfbox{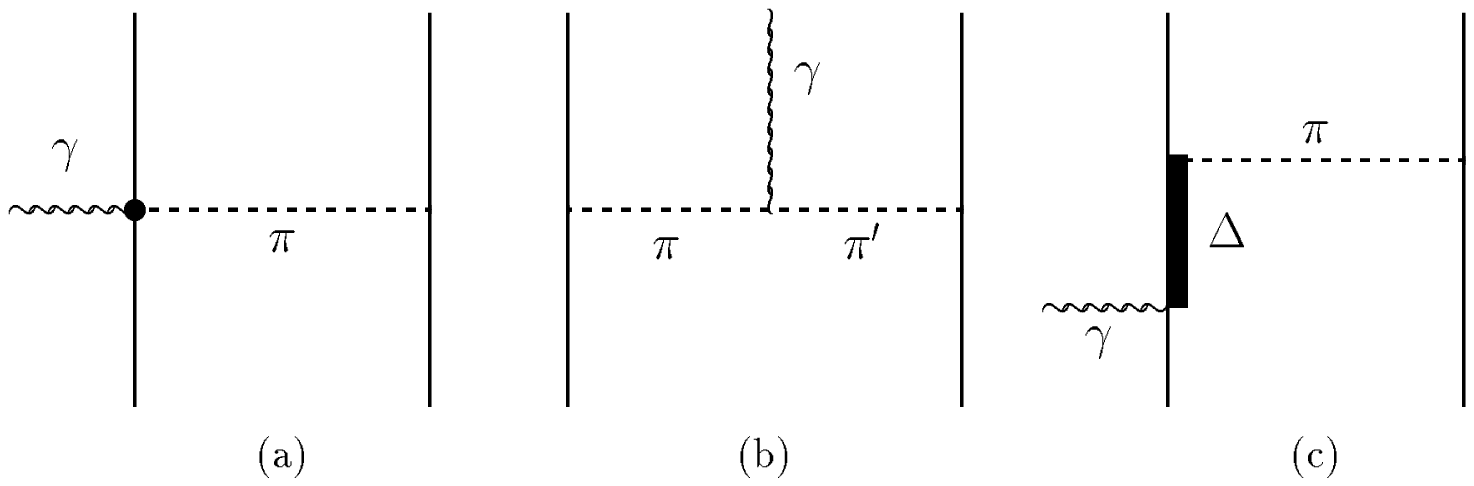}}
\end{center}
\vspace{-14cm}
\caption{\small Feynman diagrams of the three MEC operators we
consider: (a) seagull, (b) pionic and (c) $\Delta$-isobar.}
\label{feynman}
\end{figure}

We have calculated the contribution of the Meson Exchange Currents (MEC)
following the approach of refs. \cite{Gar76,Fri77,Van81}.
Using the Feynman rules we have evaluated the contributions of the
three diagrams of fig. \ref{feynman} which we call seagull (a), pionic
(b) and $\Delta$-isobar (c) terms.
%
%

For the evaluation of the MEC terms we need a model for the
pion-nucleon, photon-nucleon-$\Delta$ and pion-nucleon-$\Delta$ 
interactions. The models of the pion-nu\-cleon interaction commonly
used (pseudo-scalar and pseudo-vector couplings) produce the same
result in the non-relativistic limit. 

We define a function $h(\nr)$ as the
Fourier transform of the dynamical pion propagator
\begin{equation}
\label{hache}
h(\nr-\nr_l)=\int \frac{{\rm d}^3k}{(2\pi)^3}\,
\frac{F_{\pi\rm N}(k,\varepsilon)\, \e{i\nk\cdot(\nr-\nr_l)}}%
{k^2+m_{\pi}^2-\varepsilon^2} \, ,
\end{equation}
where we have indicated with $F_{\pi\rm N}$ the pion-nucleon form
factor and with  $\varepsilon=(\Delta E)_l$ the energy of the
exchanged pion  obtained as the difference between the energies of the
final and initial single states of the $l$-th nucleon.

In momentum space we express the seagull and pionic currents respectively 
as:
\begin{equation} 
\label{seagull}
{\bf j}^{\rm S} ({\bf q},\omega) =
4\pi\frac{f_{\pi}^2}{m_{\pi}^2} \, F_{\rm S}(q,\omega)
\sum_{k,l=1 \atop k\ne l}^A \,
[\mbox{\boldmath $\tau$}_k \times 
 \mbox{\boldmath $\tau$}_l]_{_3} \, \e{i\nq\cdot\nr_k}
 \mbox{\boldmath $\sigma$}_k  \, 
 \mbox{\boldmath $\sigma$}_l  \cdot\nabla_k h({\bf r}_k-{\bf r}_l)
\end{equation}
and
\begin{eqnarray}
\nonumber
{\bf j}^{\pi}({\bf q},\omega) & =& - 4\pi\frac{f_{\pi}^2}{m_{\pi}^2} \,
F_{\pi}(q,\omega) \,
\sum_{k,l=1 \atop k\ne l}^A \,
[\mbox{\boldmath $\tau$}_k\times 
 \mbox{\boldmath $\tau$}_l]_3 \,\,
 \int {\rm d}^3r\,\e{i\nq\cdot\nr}\, \\ \label{pionic} & &
 \mbox{\boldmath $\sigma$}_k  \cdot\nabla h({\bf r}-{\bf r}_k)\, 
 \nabla\left[
\mbox{\boldmath $\sigma$}_l \cdot\nabla h({\bf r-r}_l)
\right] \, .
\end{eqnarray}

The situation for the $\Delta$-isobar current is not so well
defined because formulations based upon static quark models
\cite{Che71,Hoc73,Loc75,Ris79} or chiral lagrangian
\cite{Pec69,Van81,Alb90} give different expressions. We have adopted 
the point of view of the first group of authors
and use the following expression:
\begin{eqnarray} 
\nonumber
{\bf j}^{\Delta}({\bf q},r) &= & -i C_{\Delta} F_{\Delta}(q,\omega) 
\sum_{k,l=1 \atop k\ne l}^A \,
\e{i\nq\cdot\nr_k}\nq \\ & & \nonumber
\times 
\displaystyle\left\{ 
[\mbox{\boldmath $\tau$}_k\times 
 \mbox{\boldmath $\tau$}_l]_3 \,\,
 \mbox{\boldmath $\sigma$}_k\times \nabla_k 
 \mbox{\boldmath $\sigma$}_l\cdot\nabla_k h({\bf r}_k-{\bf r}_l) 
\, - \, 4 \tau^3_l \nabla_k
\mbox{\boldmath $\sigma$}_l \cdot\nabla_k h({\bf r}_k-{\bf r}_l)
\right\} \, .
\label{isobar}
\end{eqnarray}
In the above equations 
$f_{\pi}=0.079$ is the effective pion-nucleon coupling constant,
$m_{\pi}$ is the pion mass and
\begin{equation}
\label{cdelta}
C_{\Delta} = 4\pi\frac{f_{\pi}^2}{m_{\pi}^2}
             \frac{4}{25M(M_{\Delta}-M)} \, ,
\end{equation}
with $M_{\Delta}$ the $\Delta$ mass. 
Finally,
$F_{\rm S}$, $F_\pi$, $F_\Delta$ show the dependence on the 
electromagnetic nucleon, delta and pion form factors. 
As we have already
discussed there are theoretical ambiguities in the choice of these
form factors. In table \ref{varmec} we show the effect of these
ambiguities on the values of the peak of the MEC responses
\cite{Ama93a}. For the seagull current, the  possible functional
dependence of the $F_{\rm S}$ form factor, generate an appreciable
uncertainty at the peak energy. For $q=550$~MeV/$c$, the uncertainty
on the maximum is $\simeq 40\%$ of its value. A similar situation is
found for the electromagnetic pion form factor $F_{\pi}$ included in
the pionic current even if the uncertainty is considerably smaller
than in the seagull case. Table \ref{varmec} shows for example that,
at $q=550$~MeV/$c$, the uncertainty of the peak value is $\sim 12\%$.

\begin{table}
\begin{center}
\begin{tabular}{|cccc|}
\hline
 $q$ [MeV/$c$]  & $\omega$ [MeV]  & 
$\Delta F_S$ [\%] & $\Delta F_\pi$ [\%]  \\ \hline
 300  & 50  & 15.7 & 5.3 \\
 400  & 90  & 25.6 & 8.3 \\
 500  & 100 & 40.2 & 12.3 \\
\hline
\end{tabular}
\caption
{\small Percentile uncertainty on the value of the peak of the 
response produced by the different choices of form factors
on seagull and pionic currents \protect\cite{Ama93a}. 
\label{varmec}}
\end{center}
\end{table}

For consistency with the one-body currents, we use in our calculations
the following expressions for $F_{\rm S}$ and $F_\pi$:
\begin{eqnarray}
F_{\rm S} & = & G_{\rm E}^{\rm P} - G_{\rm E}^{\rm N} ,\\
F_{\pi}   & = & F_{\pi\gamma} = \frac{1}{1+(q^2-\omega^2)/m_{\rho}^2}
\, ,
\end{eqnarray}
where $m_{\rho}$ is the mass of the $\rho$-meson. 

The situation is even more complicated for the $\Delta$ current since
the electromagnetic form factor $F_\Delta$ and the constant
$C_{\Delta}$ are model dependent. The major uncertainty is related to
$C_{\Delta}$, but a discussion of this problem is beyond the aim of
the present work. The expression (\ref{cdelta}) of $C_{\Delta}$ we
have chosen is widely used in the literature
\cite{Hoc73,Ris79,Ris84,Sch89}. For the form factor $F_{\Delta}$,
following the static quark model, we use:
\begin{equation}
F_{\Delta} = 2G_{\rm M}^V= G_{\rm M}^{\rm P}-G_{\rm M}^{\rm N} \, .
\end{equation}
Finally, we would like to comment on the pion-nucleon form factor
\begin{equation}
F_{\pi\rm N}(k,\varepsilon) =
\frac{\Lambda^2-m_{\pi}^2}{\Lambda^2+k^2-\varepsilon^2}
\end{equation}
present in the pion propagator of eq. (\ref{hache}). We have verified
\cite{Ama93a,Ama93b} that, in the QE peak region,  for the values of
$\Lambda$ commonly accepted ($\sim 1$~GeV), the results are very close
to those obtained considering simply $F_{\pi\rm N} =1$, which is the
value we have adopted.

The MEC we have presented above act only on the transverse
response. In principle MEC contributions are present also in the
longitudinal response, but it has been shown \cite{Lal97} that their
effects are very small, of the order of 3\% at the energy of the
peak. For this reason in the following we shall neglect the
corrections to the longitudinal response and we shall discuss only the
effect of MEC on the transverse one.

The MEC are two-body operators, therefore they can produce the
excitation of 1p-1h or 2p-2h pairs. In the excitation of 1p-1h pairs,
the transition amplitudes of MEC and those of one-body currents
interfere. These interference terms produce the larger contribution of
the MEC to the 1p-1h response.
%
%
\begin{figure} 
\vspace{-4cm}
\begin{center}                                                                
\leavevmode
\epsfysize = 500pt
\hspace*{0.45cm}
\makebox[0cm]{\epsfbox{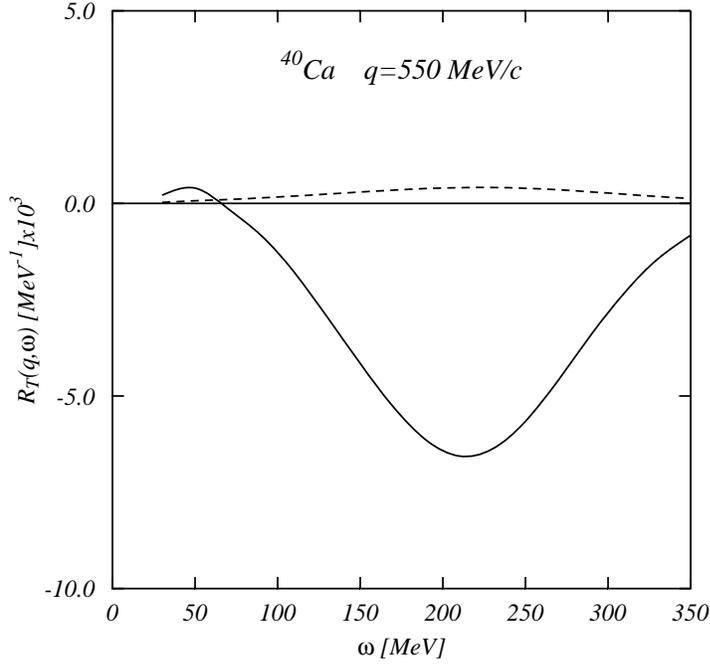}}
\end{center}
\vspace{-4.5cm}
\caption{\small 1p-1h MEC contributions to the excitation of the
$^{40}$Ca transverse response at $q$=550~MeV/$c$. The full line shows
the interference term between one-body current and MEC. The dashed
line represents the pure MEC contribution.
}
\label{interference}
\end{figure}
%
%
%

%
%
\begin{figure}
\begin{center}                                                                
\leavevmode
\epsfysize = 500pt
\hspace*{-0.5cm}
\makebox[0cm]{\epsfbox{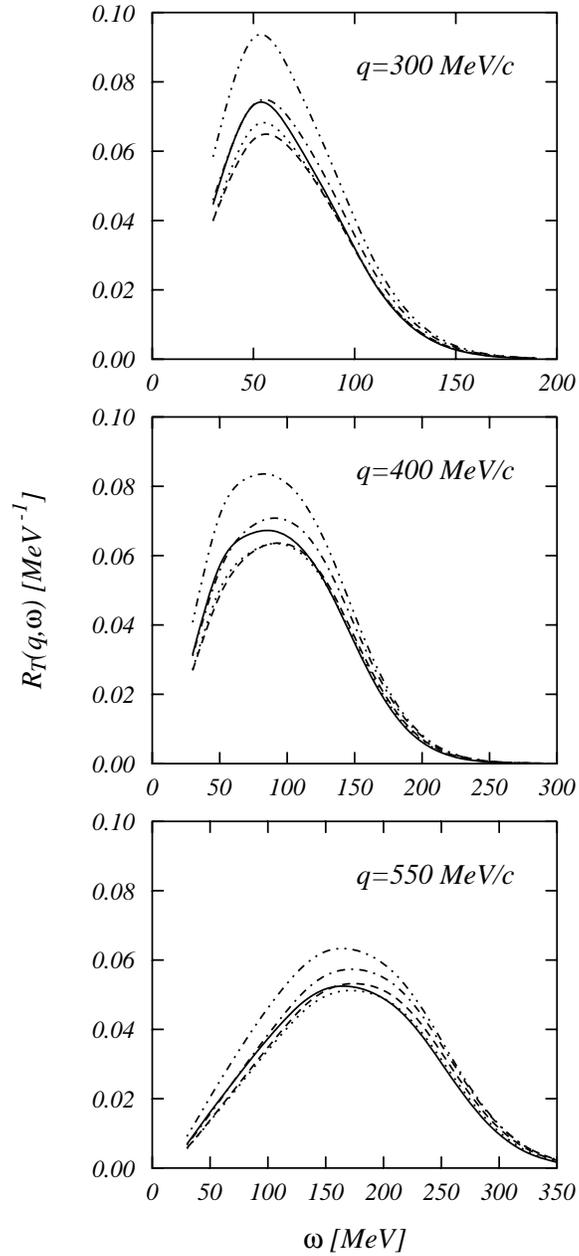}}
\end{center}
\vspace{-1.cm}
\caption{\small 1p-1h transverse responses of $^{40}$Ca.
The dashed-dotted lines are the one-body responses. The remaining lines
show the results obtained by adding the various parts of MEC:
seagull (dashed-doubly-dotted), pionic (dashed), $\Delta$ isobar
(dotted), seagull+pionic+$\Delta$ isobar (solid). 
}
\label{1p1hresponse}
\end{figure}

In fig. \ref{interference} we show the contribution of the MEC to
the transverse response of $^{40}$Ca for $q$=550~MeV/$c$. While the
contributions of pure MEC are positive for all excitation energies,
the contribution of the interference term between one-body current
and MEC is positive at small energies and become negative at energies
above 80~MeV.  
%
%
\begin{figure}[ht]
\vspace{-7cm}
\begin{center}                                                                
\leavevmode
\epsfysize = 500pt
\hspace*{0.05cm}
\makebox[0cm]{\epsfbox{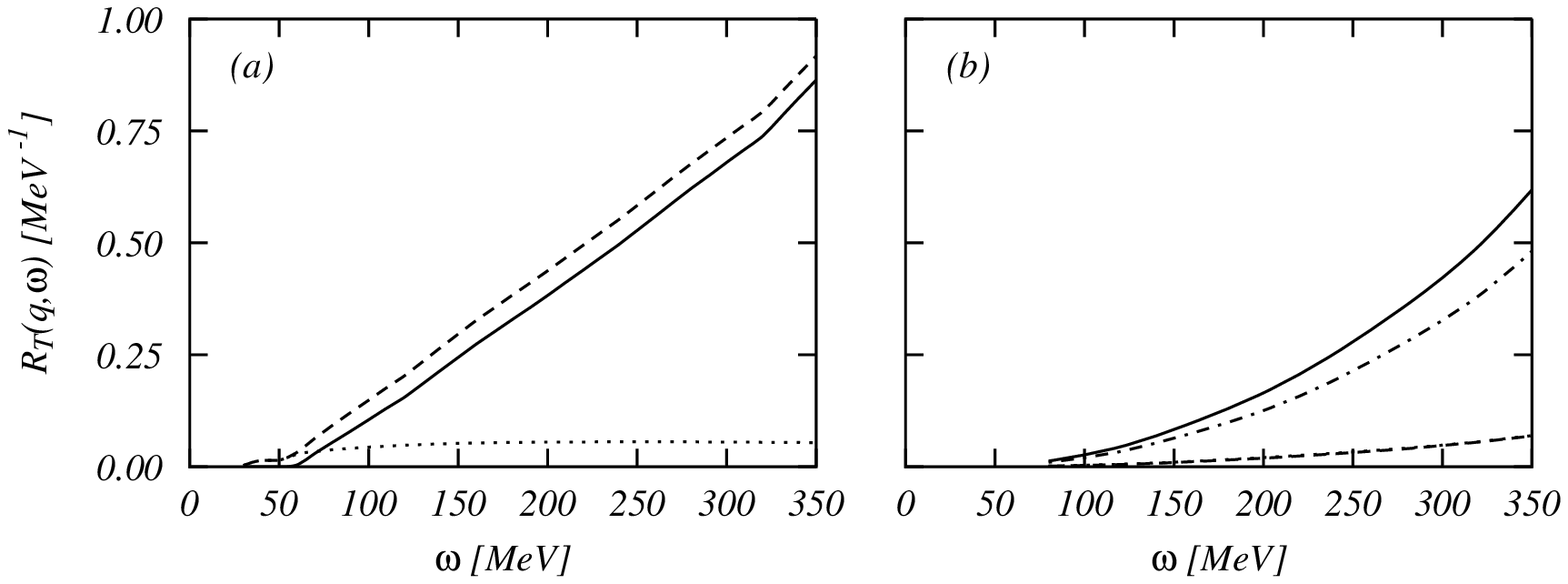}}
\end{center}
\vspace{-4.5cm}
\caption{\small 2p-2h responses of $^{12}$C for $q=550$~MeV/$c$. 
Panel (a): Seagull plus pionic contributions. 
The dashed line is the term corresponding to the two particle emission
while the dotted line has been obtained for the case of one particle
in the continuum and the other one on an excited bound state. The full 
curve represents the total 2p-2h response. Panel (b): Two nucleon
emission contributions induced by the $\Delta$ isobar current. The 
dashed-dotted line corresponds to the emission of a proton-neutron
pair. The dotted and dashed lines give the contribution of the
two-proton and two-neutron emission, respectively, and are overlapping
in the scale of the figure. The solid line is the total response.
}
\label{2p2h}
\end{figure}
%
%
%

%
%
\begin{figure}
\begin{center}                                                                
\leavevmode
\epsfysize = 500pt
\hspace*{-0.7cm}
\makebox[0cm]{\epsfbox{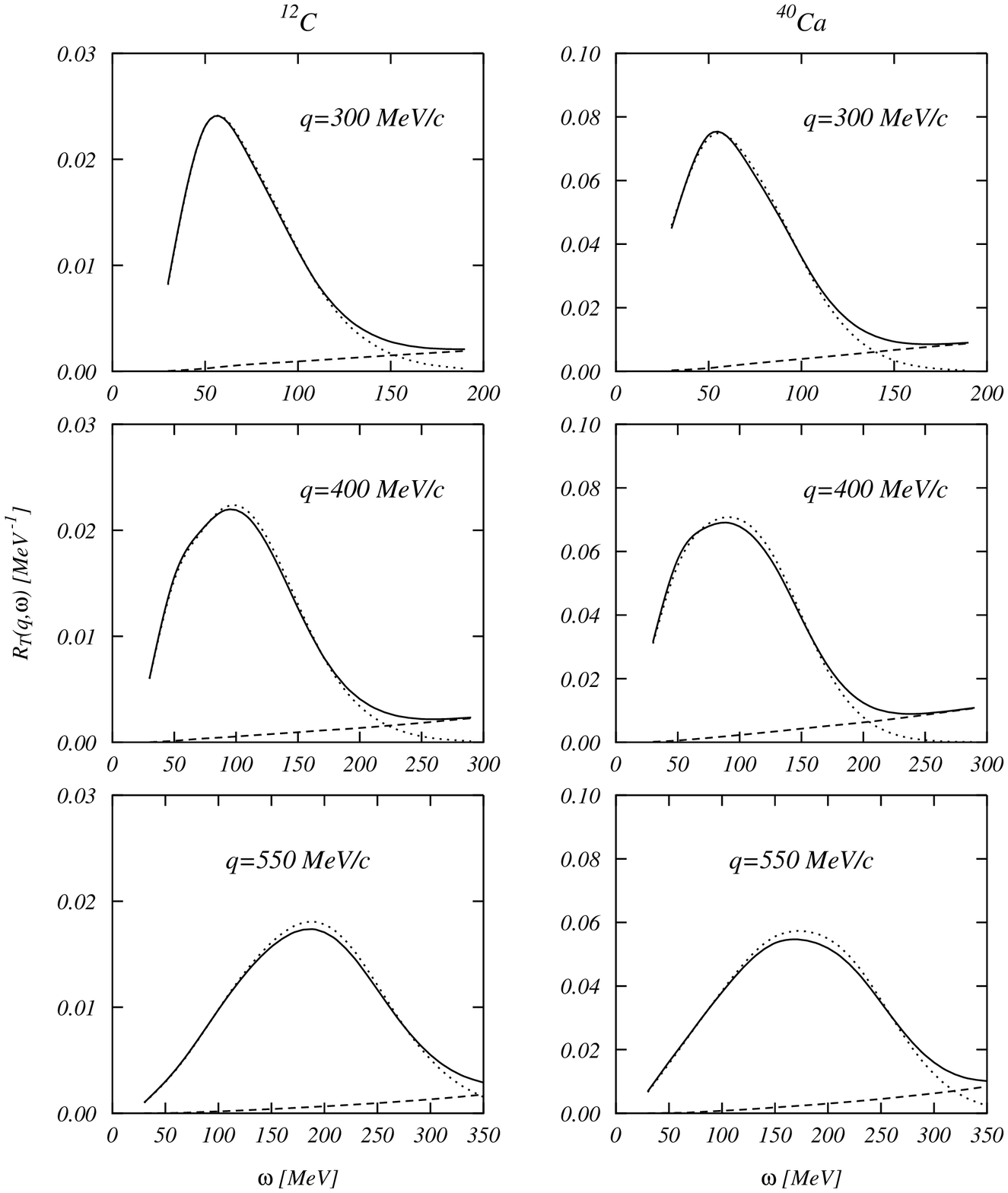}}
\end{center}
\vspace{-1.cm}
\caption{\small Transverse responses calculated in $^{12}$C and
$^{40}$Ca. The solid lines represent the total responses including
1p-1h and 2p-2h final states. The dotted lines give the one-body
response and the dashed lines the 2p-2h contributions. The
dashed-dotted lines have been obtained with one-body operators only.
}
\label{fullresp}
\end{figure}

In agreement with \cite{Koh81} we find that the effect of the pionic
and seagull currents on the responses have opposite sign. 
The pionic
current diminishes the pure one-body response, while the seagull
increases it. In fig. \ref{1p1hresponse} we show this fact for the
$^{40}$Ca nucleus, but we found analogous results also for the
$^{12}$C response \cite{Ama93a}. Considering only seagull and pionic
currents we obtain an increase of the transverse response in the peak
region \cite{Ama92,Ama93b}. The inclusion of the $\Delta$ isobar current
gives a negative contribution. The final results consists in a
lowering of the one-body responses. The effect of the seagull and
pionic currents becomes smaller as $q$ increases while that of the
$\Delta$ isobar current remains roughly constant. For this reason, the
total lowering effect of the MEC increases its magnitude with $q$.

In all the nuclear models we have used to calculate the response,
the nuclear 1p-1h final states are orthogonal to the 2p-2h final
states. This means that in our approach there is not interference
between 1p-1h and 2p-2h excitation amplitudes and, as a consequence,
the 2p-2h responses add to the 1p-1h ones. 

The two body MEC operators can lead to different kinds of 2p-2h final
states. A first classification concerns the number of particles which
are in the continuum. A first type of 2p-2h state is produced when
none of the two particles is in the continuum. 
These are low energy excitation and they do not
contribute in the QE region. A second kind of excited state is formed
when one particle is in the continuum and the other one is lying on a
bound single particle state.
Finally the last type of excited 2p-2h state is the one which has
both particles in the continuum. 

In the panel (a) of fig. \ref{2p2h} we show the behavior of the 2p-2h
responses for the various types of final states. In this calculation
only the seagull and pionic currents have been used. The contribution
of the response where only one particle is in the continuum rapidly
reaches  a plateau value and it remains roughly constant with the
increase of the excitation energy. Since the number of particle-hole
configurations entering in the excitation of this part of the response
remains constant even if the energy increases. The two-nucleon
emission contribution behaves quite differently: its value increases
steadily with the value of the excitation energy. This is related to
the enlarging of the phase space available for the final states of
this reaction. 

The other possible classification of 2p-2h final states is related to
the kind of particles which are emitted. Seagull and pionic currents
are generated by the exchange of charged pions. The $\Delta$
isobar current is active also when a neutral pion is exchanged. Then,
while the seagull and pionic currents can lead to final 2p-2h states
where only a proton and a neutron pair is emitted, the $\Delta$ isobar
current can emit also a pair of like nucleons. 

The contribution of these different kind of excited states on the
2p-2h response of the $\Delta$ isobar current is shown in
panel (b) of fig. \ref{2p2h}. The two-proton and two-neutron emission
responses are indistinguishable at the scale of the figure. This is
due to the fact that $^{12}$C nucleus is practically symmetric in
terms of proton-neutron single-particle structure. The number of
protons and neutrons is the same, and the small differences in the
single-particle energies and in the wave functions affect very little
the response. The dominant emission channel remains the
proton-neutron one, and its relative contribution becomes more
important with increasing energies.

In fig. \ref{fullresp} we present the total QE transverse responses
for $^{12}$C and $^{40}$Ca. The 2p-2h responses are rather small below
the peak but become important at higher energies.

A detailed analysis of the MEC contribution \cite{Ama94a} shows that
the pionic and seagull currents produce an increase of the values of
the responses in the peak. The inclusion of the $\Delta$ isobar current
has an opposite effect. If for $q$=300~MeV/$c$ the total responses are
similar to the one-body responses, for higher values of $q$ they are
even smaller. There is not big difference between the two nuclei,
though in $^{40}$Ca the modifications are slightly bigger. The greater
effect we found is less than 5\%, certainly within the range of
uncertainty related to the choice of the nucleon and mesons form factors.

\section{The nuclear excitation}

In this section we shall present some of the models commonly used to 
describe the nuclear QE excitation. In this region the excitation
energy of the nucleus is well above the continuum threshold, therefore
at least one particle is emitted from the nucleus. Due to the
relatively high values of energy and momentum transfer the QE
excitation is not of collective type like vibrations or rotations, but
the process is dominated by the single particle dynamics. The MF model
is the obvious starting point of any description of this
excitation. We shall first present the basic ideas of the MF model and
in subsequent subsections we shall discuss the validity of the
approximations made in this treatment. 

\subsection{Mean field models}
\label{mfmodel}
In the MF model the many-body nuclear states are Slater determinants
of sp wave functions. In the ground state all the sp states below the
Fermi level are occupied and all the states above it are empty. The
excited states of the system are generated by promoting nucleons from
below to above the Fermi level. These basic ideas of the MF approach
can be used with different set of sp bases, and this differentiate the
various MF models. In our presentation we shall treat the shell
model where the sp basis is generated considering the finite size of
the nucleus and the FG model where the system is considered to be infinite
and translationally invariant.

\subsubsection{The Continuum Shell Model}
\label{csm}
The basis commonly used to describe finite nuclear systems is composed
of sp states characterized by the orbital angular momentum $l$, the
total angular momentum $j$, its projection $m$ on the $z$ axis and the
isospin $z$ axis projection, $t$. For discrete sp states it is also necessary
to specify the principal quantum number $n$, 
\begin{equation} 
\label{spwave_dis}
|nljmt \rangle \equiv \psi^t_{nljm}(\nr) 
\equiv R^t_{nlj}(r) \, {\cal Y}_{ljm}(\hat{r}) \, \chi_t \,
\end{equation}
where
\begin{equation} 
{\cal Y}_{ljm}(\hat{r}) \, = \, \sum_{\mu s} \, 
\langle l \mu \half s | j m \rangle \, 
Y_{l \mu}( \hat{r} ) \, \chi_s \, .
\end{equation}
Here $\langle l \mu \half s | j m \rangle$ is a Clebsh-Gordan
coefficient, $\chi_s$ is the spin function, with $s$ the spin $z$ axis
projection, and $\chi_t$ is the isospin function. For continuum sp 
states their energy $\epsilon$ must be given
\begin{equation} 
\label{spwave_con}
|\epsilon ljmt \rangle \equiv \psi^t_{ljm}(\nr,\epsilon) 
\equiv R^t_{lj}(r,\epsilon) \, {\cal Y}_{ljm} \, \chi_t \, .
\end{equation}

One possibility of generating the set of sp states is to use a
Hartree-Fock procedure as is done in ref. \cite{Koh83}. This
implies the choice of an effective nucleon-nucleon interaction and the
solution of a system of non linear Schr\"odinger-like differential
equations.

We use a different approach called Continuum Shell Model (CSM). We fix
a central real potential $V(r)$ acting on all the nucleons, and we
generate the sp states (\ref {spwave_dis}) and (\ref {spwave_con}) by 
solving the traditional radial sp Schr\"odinger equation with the
appropriate boundary conditions on the radial parts. The solutions of
this equation should be regular at the origin. At high $r$ values
bound states have exponential decay, while continuum states have
oscillating behavior. 
%
%
\begin{figure}
\begin{center}                                                                
\leavevmode
\epsfysize = 500pt
\hspace*{-0.85cm}
\makebox[0cm]{\epsfbox{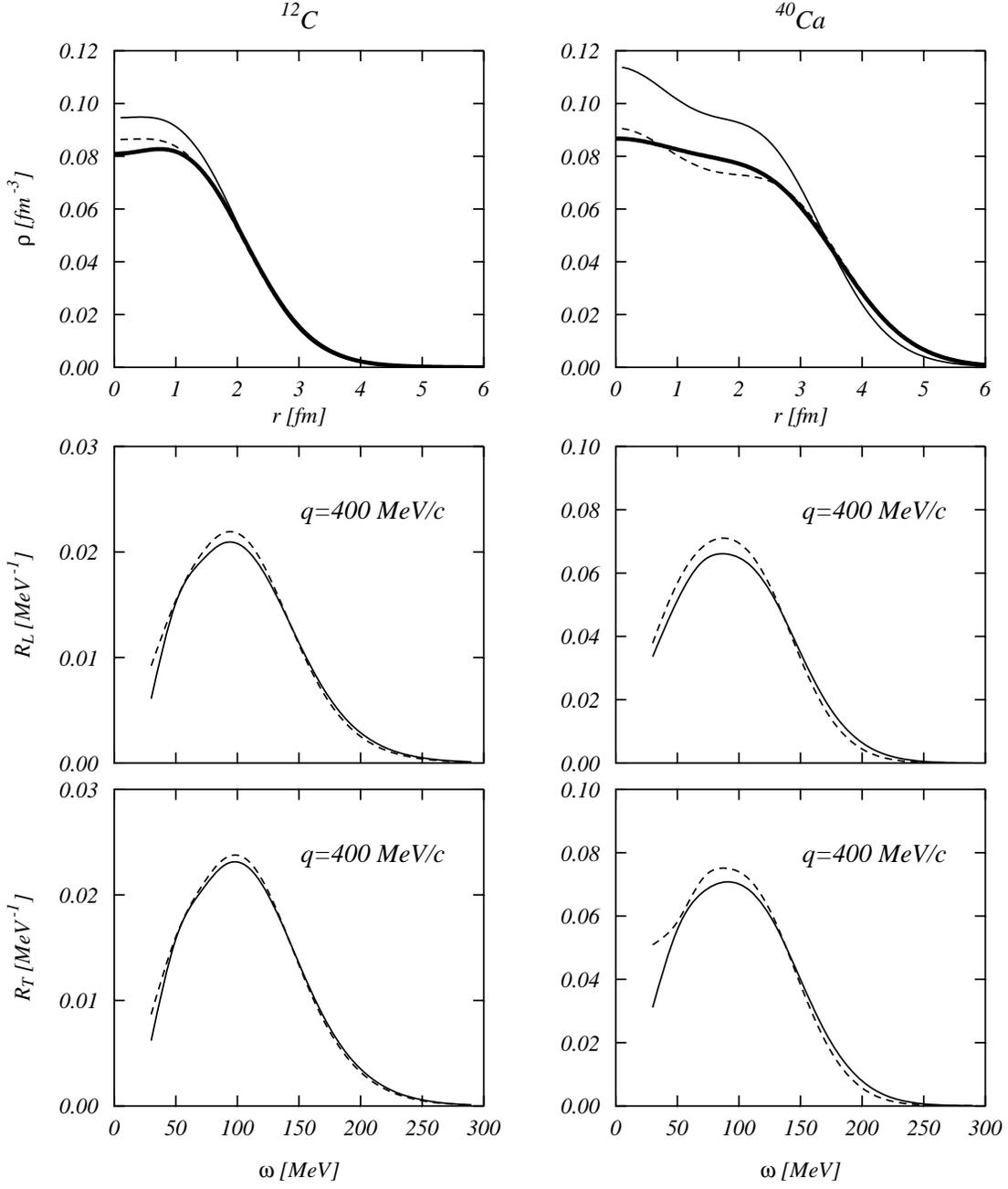}}
\end{center}
\vspace{-1cm}
\caption{\small Sensitivity of the responses to the choice of the
single particle wave functions. Full lines represent the results
obtained with the Woods-Saxon set of
refs. \protect\cite{Co'84,Co'85,Co'87b}. Dashed lines correspond to
a new parameterization aiming to reproduce the charge density
distributions. Thick full lines represents the empirical charge
density distributions reconstructed using the Fourier-Bessel
parameters given in ref. \protect\cite{DeJ87}.
}
\label{woods}
\end{figure}

As we deal with one- and two-body operators, the only possible final 
states are of 1p-1h and 2p-2h types, with the particles in the
continuum, which in this model are, respectively, given by: 
\begin{equation}
\label{1p-1h-fstat}
| \beta,JM_J \rangle \, = \, 
| (\epsilon_p l_p j_p),(n_h l_h j_h);JM_J  \rangle \, ,
\end{equation}
\begin{equation}
\label{2p-2h-fstat}
| \beta,JM_J \rangle \, = \, 
| \left[ (\epsilon_{p_1} l_{p_1} j_{p_1}),
         (\epsilon_{p_1} l_{p_2} j_{p_2});J_p \right], 
\left[ (n_{h_1} l_{h_1} j_{h_1}),
         (n_{h_1} l_{h_2} j_{h_2});J_h \right]; JM_J  \rangle \, . 
\end{equation}

In the case of one-body operators, only the 1p-1h final states can be
excited. The sum over $(\beta J)$ in eqs. (\ref{rl1}) and (\ref{rt1})
goes on all the possible 1p-1h configurations allowed by the
angular momentum and parity composition rules. This means that
$|j_p-j_h| \leq J \le j_p+j_h$ and $(-1)^{l_p+l_h}$ should be equal to
the parity of the corresponding electromagnetic transition. The 
transition amplitudes can be written as:
\begin{equation}
\label{obtrans}
\langle \beta,J \parallel O^{\rm OB}_J(q) \parallel 0 \rangle \, = \,
\langle \epsilon_p l_p j_p \parallel O^{OB}_J(q) \parallel n_h l_h j_h
\rangle \, 
\delta(\omega-\epsilon_p+\epsilon_h) \, .
\end{equation}

In case of two-body operators the expressions are slightly more
complicated, and it is possible to excite both  1p-1h and 2p-2h final
states. In the first case the same angular momentum, parity and energy
rules as before apply and the transition amplitudes can be written as:
\[
\langle \beta,J \parallel O^{\rm TB}_J(q) \parallel 0 \rangle \, = \, 
\sum_{\alpha <  \epsilon_F} \, \sum_{J_{1} J_{2}} \, 
(-1)^{J_1+J_2+J+1} \, \sqrt{(2J_1+1)(2J_2+1)} \, 
\sixj{J_1}{J_2}{J}{j_h}{j_p}{j_\alpha} 
\]
\vspace*{-.7cm}
\begin{eqnarray}
\nonumber
\label{tbtrans1}
& &\hspace{-0cm} \left[ \, 
\langle (n_\alpha l_\alpha j_\alpha), (\epsilon_p l_p j_p); J_1 
\parallel O^{\rm TB}_J(q) \parallel 
(n_h l_h j_h),(n_\alpha l_\alpha j_\alpha); J_2 \rangle \, + \right. \\ 
\nonumber
& & \hspace{0.1cm} \left. (-1)^{j_\alpha+j_h+J_2+1} \,
\langle (n_\alpha l_\alpha j_\alpha), (\epsilon_p l_p j_p); J_1 
\parallel O^{\rm TB}_J(q) \parallel 
(n_\alpha l_\alpha j_\alpha),(n_h l_h j_h); J_2 \rangle \, \right] \, \\
& & \delta(\omega-\epsilon_p+\epsilon_h) \, ,
\end{eqnarray}
where we have used a 6j coefficient.

If the final state has two particles in the continuum the transition
amplitude is: 
\[
\hspace*{-3.3cm}
\langle \beta,J \parallel O^{\rm TB}_J(q) \parallel 0 \rangle \, = \, 
(-1)^{j_{p_1}+j_{p_2}+J_p} \, 
\delta (\omega - \epsilon_{p_1} - 
\epsilon_{p_2} + \epsilon_{h_1} + \epsilon_{h_2} )
\]
\vspace*{-.7cm}
\begin{eqnarray}
\label{tbtrans2}
& &\hspace{-.55cm} \left[ \, 
\langle (\epsilon_{p_2} l_{p_2} j_{p_2}), 
(\epsilon_{p_1} l_{p_1} j_{p_1}); J_p 
\parallel O^{\rm TB}_J(q) \parallel 
(n_{h_1} l_{h_1} j_{h_1}),
(n_{h_2} l_{h_2} j_{h_2}); J_h \rangle \, + \right. \\ 
\nonumber
& & \hspace{-0.5cm} \left. (-1)^{j_{h_1}+j_{h_2}+J_h+1} \,
\langle (\epsilon_{p_2} l_{p_2} j_{p_2}), 
(\epsilon_{p_1} l_{p_1} j_{p_1}); J_p 
\parallel O^{\rm TB}_J(q) \parallel 
(n_{h_2} l_{h_2} j_{h_2}),
(n_{h_1} l_{h_1} j_{h_1}); J_h \rangle \, \right] \, ,
\end{eqnarray}
where now $|J_p-J_h| \leq J \le J_p+J_h$.

The above expressions show that in the MF model, the complicated
problem of evaluating many-body transition matrix elements
has been reduced to the much simple problem of calculating
sums of one- and two-body transition matrix elements. 
The calculation of the sp transition matrix elements is long and
tedious, but it does not present fundamental difficulties and it can
be carried on without any further hypothesis or approximation.
The explicit expressions of the matrix elements can be found in 
ref.~\cite{Ama93a}. 

We have studied the sensitivity of the CSM results to the choice of the
potential parameters \cite{Ama93a,Ama93b}, the only input of the
model. We have performed CSM calculations with two sets of
parameters. The first one taken from the literature has been fixed to
reproduce the sp energies around the Fermi level and it has been used
to describe low-lying and giant resonance excitations
\cite{Co'84,Co'85,Co'87b}. The second set of parameters has been fixed
to have good description of the empirical charge density distribution
\cite{Ama93a,Ama93b}. The comparison of the results obtained for
one-body responses is shown in fig. \ref{woods}. The difference
between the results obtained with the two parametrizations becomes
smaller at high values of $q$ and $\omega$. This indicates that in the
QE peak the surface effects are not important. The differences between
the two results are  compatible with the uncertainty related to the
choice of the nucleon form factors. All the CSM results we shall
present have been obtained with the set of parameters of
refs. \cite{Co'84,Co'85,Co'87b}. 

\subsubsection{Fermi gas  model}
\label{fg}
An alternative MF model often used 
\cite{Czy63,Mon69,Koh81,Van81,Bou91,Gar92,Gil97}
to describe the QE excitation is the FG model. In this model the
nucleus is treated as an infinite system of non-interacting
fermions. Both the volume $\Omega$ of the system and number of
particles $A$ are infinite, but the density $\rho=\Omega/A$ remains
finite. In this  model the  meaningful quantities are the responses
per nucleon.

Since the nuclear system is translationally invariant, the adequate
set of sp wave functions is formed by eigenstates of the momentum
\begin{equation} 
\label{fgwave}
|kst \rangle \equiv \psi^t_{ks}(\nr) \equiv 
\sqrt{\frac{1}{\Omega}} \,\,\,  e^{i\nk \cdot \nr} \chi_s \chi_t \, .
\end{equation}
In the nuclear ground state all the sp states with $k<k_{\rm F}$ are
occupied. The Fermi momentum $k_{\rm F}$ is related to the constant
density of the system by the expression:
\begin{equation} 
\label{kfdens}
k_{\rm F}\, =\, \left( \frac{3}{2}\pi^2 \rho \right)^{1/3} \, .
\end{equation}
from which, assuming a hard sphere model of the nucleus,
$\rho=\left( \frac{4}{3} \pi r^3_0\right)^{-1}$ and 
for $r_0$ = 1.12~fm, one obtains the well known values
$k_{\rm F}$ = 1.36~fm$^{-1}$ or 268~MeV/$c$.

In analogy with the CSM the excited states are generated by promoting a
particle from a level below to above the Fermi level. While in the CSM
model the angular momentum of the system should be conserved, in the
FG model the total momentum of the system is conserved. This means that the
relation $\nq+\np_i=\np_f$, where $\np_i$ and $\np_f$ are respectively
the initial and final nucleon momenta, should always be satisfied. 
For the Pauli exclusion principle $|\np_f|$ should be greater than 
$k_{\rm F}$, and therefore those excitations such that
$|\nq +\np_i|<k_{\rm F}$ are forbidden. If 
$|\nq|>2k_{\rm F}$ there are no forbidden excitations, i.e. every
nucleon can be emitted to the continuum.

In the FG model there are not discrete sp states, therefore it cannot
be used to study low-lying nuclear excited states. Furthermore the
excitation spectrum of this model does not predict the giant
resonances, since they are collective surface vibrations of the
nucleus in the continuum and the FG does not have surface. The QE
region is however well suited to be studied with the FG model. The
excitation energy is such that the nuclear finite state is always in
the continuum and the nuclear dynamics is of single-particle type
which one expects to be more related to the volume than surface effects.

The expression (\ref{fgwave}) of the sp states allows one to perform
a large 
part of the calculation of the responses in an analytic way. The
final expressions for the responses to be handled numerically are much
simpler than those obtained in the CSM. These expressions can be found
in \cite{Ama93a,Ama94b}.

The shape of $R_L$ and $R_T$ changes when $q$ become larger that
$2k_{\rm F}$. For $q<2k_{\rm F}$ the response as a function of the
excitation energy $\omega$  has first a linear increase then it
decreases as a parabola. For $q>2k_{\rm F}$ the response has a
parabolic behavior with a maximum in $\omega= q^2/2M$. The energy and
momentum conservation implies that for a fixed value of $q$ there is a
maximum value of $\omega$ where the response is different from zero.
Of course the experimental data do not show these behaviors.

\begin{figure}  
\begin{center}                                                                
\leavevmode
\epsfysize = 500pt
\hspace{-0.85cm}
\makebox[0cm]{\epsfbox{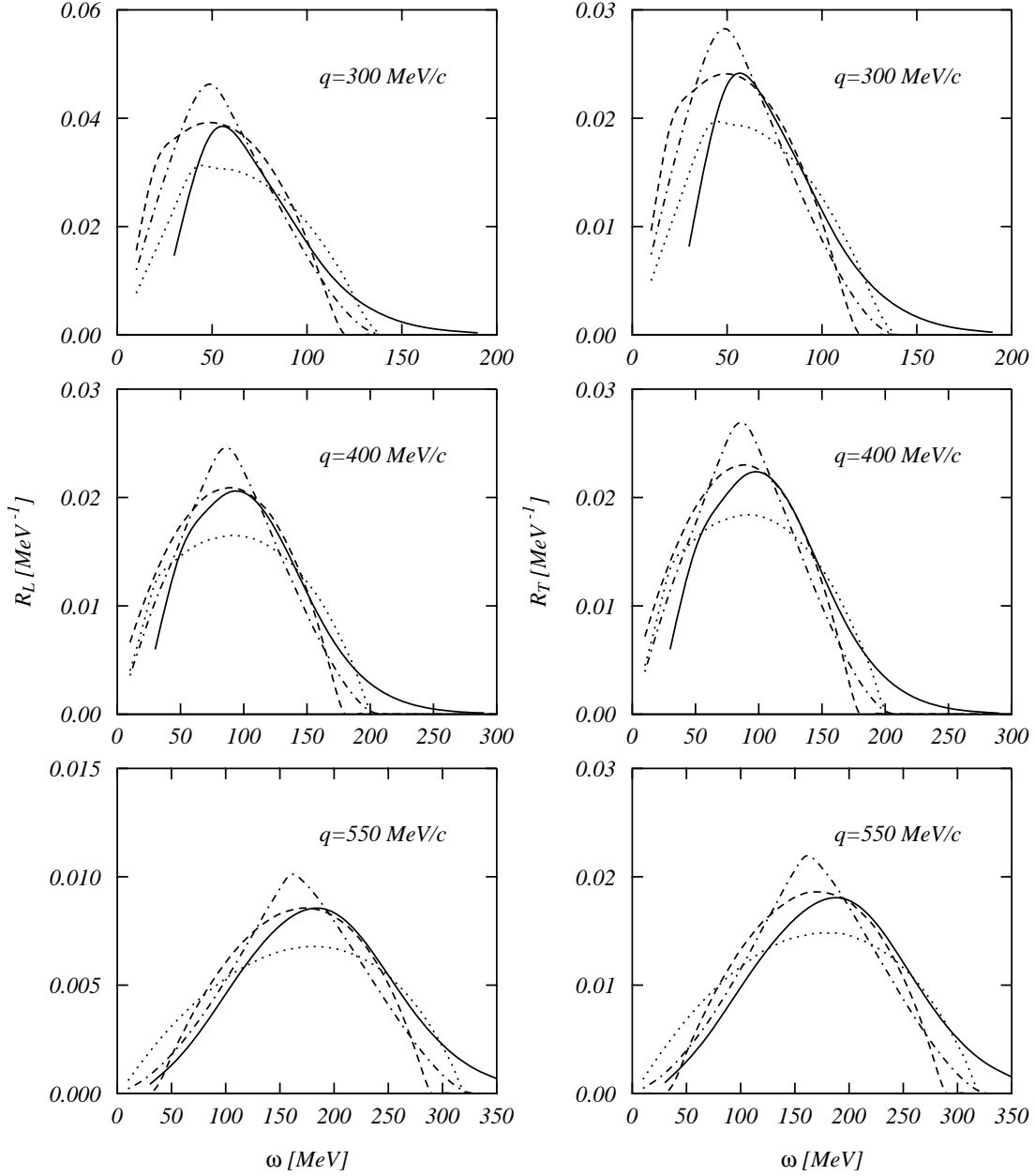}}
\end{center}
\vspace{-1.cm}
\caption{\small Comparison between shell model responses in $^{12}$C
(full lines) and FG responses. The dotted lines have been calculated 
using the Fermi momentum value extracted from the nuclear central
density $k_{\rm F}$=268~MeV/$c$, the dashed lines with $k_{\rm
F}$=215~MeV/$c$ calculated with eq. (\ref{kav}) and the dashed-dotted
lines with the LDA, eq. (\ref{lda}).
}
\label{fglda}
\end{figure}

We have studied the reliability of this model in the description of the
QE excitation by comparing the nuclear responses obtained with CSM
calculations with those obtained with the FG model. The only input in
the FG calculations is the value of $k_{\rm F}$, therefore of the
nuclear density. The results we have obtained for the traditional
nuclear matter value of $k_{\rm F}$, are represented by the dotted
lines of fig. \ref{fglda}. These responses are quite different
from those calculated within the CSM (full lines). The maxima are
smaller and the widths are broader. It is remarkable that the positions
of the peaks are roughly the same for both calculations.

%
%
%
%
%

%
%
\begin{figure}
\begin{center}                                                                
\leavevmode
\epsfysize = 500pt
\hspace{-0.55cm}
\makebox[0cm]{\epsfbox{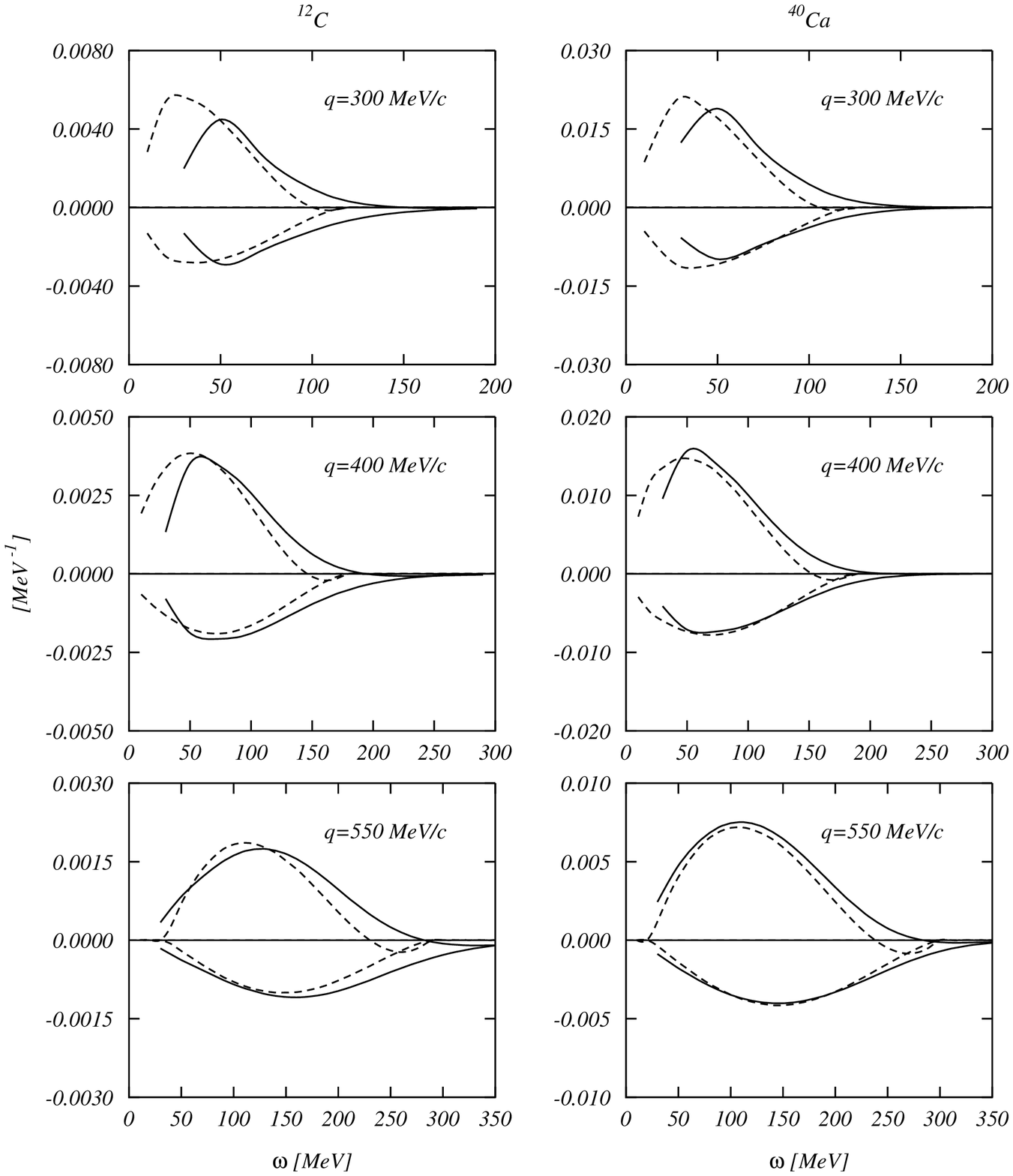}}
\end{center}
\vspace{-1.cm}
\caption{\small Interference terms between pionic and seagull MEC
and one-body currents for $12$C (left panels)  and $40$Ca (right
panels). The full lines have been calculated within the shell model,
while the dashed lines show the FG results obtained using the average
values $k_{\rm F}$=215~MeV/$c$ and $k_{\rm F}$=235~MeV/$c$ for the two
nuclei, respectively. The pionic contribution is negative and that of
the seagull current positive.
}
\label{fgmec}
\end{figure}

The value of $k_{\rm F}$ used in these calculations has been obtained
under the assumption that the density of the FG system corresponds to
the central density of the nucleus. A better approximation consists in
considering an average nuclear density  by using a value of the Fermi
momentum evaluated as:
\begin{equation} 
\label{kav}
\langle k_{\rm F} \rangle \, = \, \displaystyle 
\left( \frac{3}{2}\pi^2 \right)^{1/3} \, 
\frac{ \displaystyle 
\int \, {\rm d}r \, r^2 \, \left[ \rho(r) \right]^{4/3} }
     { \displaystyle \int \, {\rm d}r \, r^2 \, \rho(r) }
\end{equation}
In our calculations the effective momenta for $^{12}$C and $^{40}$Ca
have been obtain using the proton density produced by the
Woods-Saxon potential (see fig. \ref{woods}). We found 
$\langle k_{\rm F} \rangle$=215~MeV/$c$ for $^{12}$C and 
$\langle k_{\rm F} \rangle$=235~MeV/$c$ for $^{40}$Ca. The use of the 
effective Fermi momentum greatly improves the agreement with the CSM
results as is shown in fig. \ref{fglda}. As expected the agreement
is better the heavier the nucleus is and the higher the momentum
transfer becomes. This is a further indication that, under these
conditions, the QE excitation is mainly a volume effect and the
nuclear surface plays a minor role.

The Local Density Approximation (LDA) is a prescription commonly used
\cite{Gil97} to correct the FG calculations for the fact that the
nuclear density is not constant. The philosophy underlying the LDA is
that, at any point of the finite nuclear volume, the response can be
described by the nuclear matter response calculated at the
corresponding density. Using eq. (\ref{kfdens}) we evaluated the LDA
responses as:
\begin{eqnarray} 
\nonumber
R^{\rm LDA}(q,\omega) &\equiv&
\int \, {\rm d}^3r \, \frac{ R^{\rm FG}(q,\omega,k_{\rm F}(r))} 
{\Omega(r)} \\
&=& \frac{2}{3 \pi^2 A}
\int \, {\rm d}^3r \, k^3(r) \, R^{\rm FG}(q,\omega,k_{\rm F}(r))
\label{lda}
\end{eqnarray}
where we have indicated with $R^{\rm FG}$ the FG response, and with
$\Omega(r)$ the nuclear volume defined as $\Omega(r)= A/ \rho(r)$.

The dashed-dotted lines of fig. \ref{fglda} are the LDA responses
calculated using the shell model proton density. The shapes of the LDA
responses are more similar to those of the CSM than those obtained
with the FG. While these latter have a sharp cut in energy, the
LDA produces a tail, as in the CSM calculations. In spite of the
qualitative similarity between LDA and CSM responses, from the
quantitative point of view the agreement between these two results is
rather poor.

We have discussed so far the FG results obtained for one-body
operators only. We have also calculated the MEC contribution in this
model, by considering the most important terms, i.e. those produced
by the interference with the one-body currents. In fig. \ref{fgmec}
we compare
the interference terms of pionic and seagull currents with the
one-body currents calculated in the CSM and in the FG model (using effective
momenta). Also in this case the FG gas model produces
results very similar to those obtained in CSM, especially for high
values of $q$.

\subsection{Relativistic corrections}
\label{relstr}
In section \ref{relem} we discussed the relativistic corrections
to the expressions of the one-body electromagnetic operators. The
results presented there have been obtained using the relativistic 
energy-momentum relation $E^2-p^2 = M^2$. The CSM and the FG model are
non relativistic, therefore the use of the expressions 
(\ref{relcharge}) and  (\ref{relcurrent}) is not correct within these 
models.
%
%
\begin{figure}
\begin{center}                                                                
\leavevmode
\epsfysize = 500pt
\hspace*{-0.85cm}
\makebox[0cm]{\epsfbox{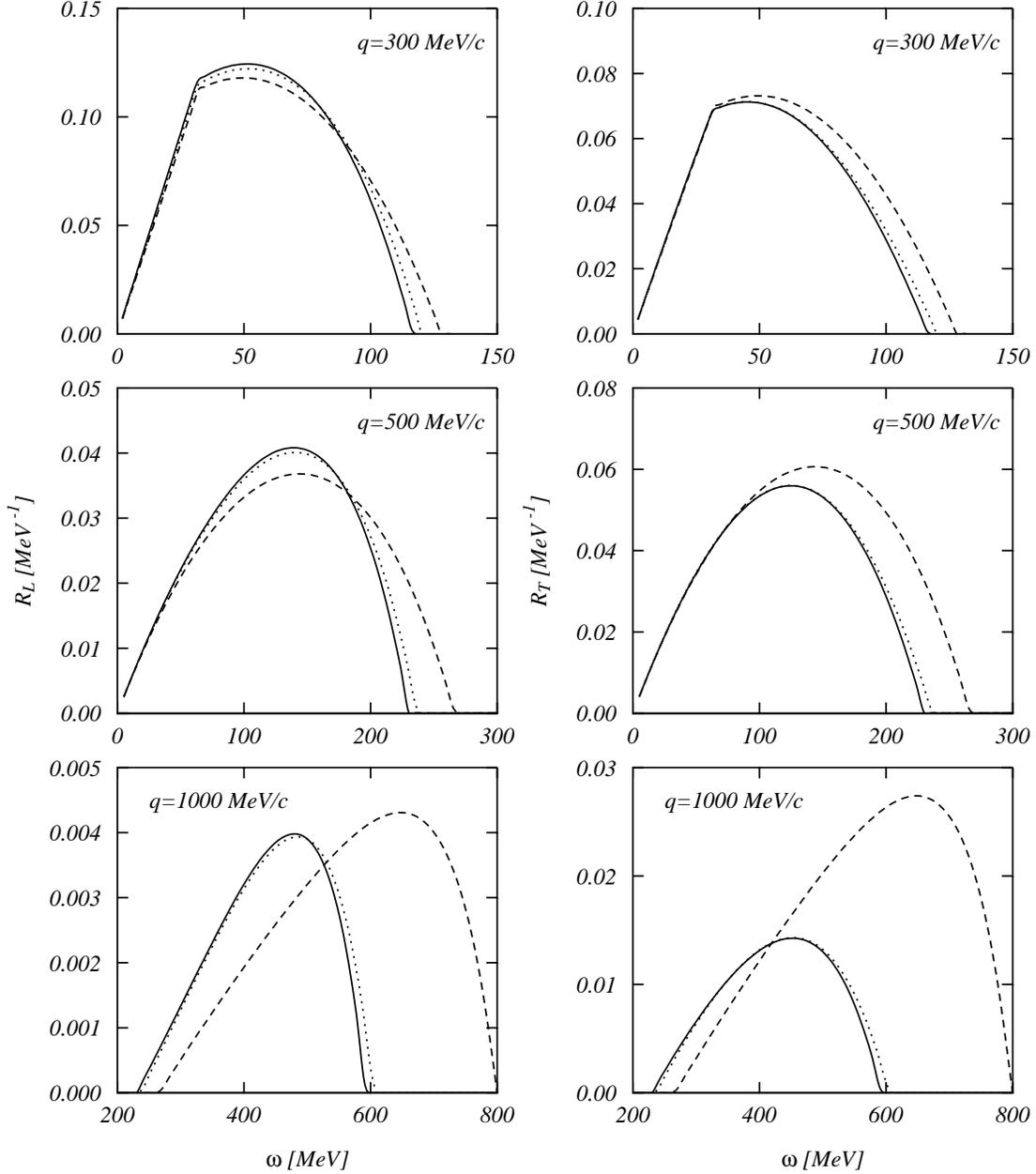}}
\end{center}
\vspace{-1cm}
\caption[]{\small The full lines show the results of the 
exact relativistic FG model. The dotted lines have been obtained
considering the relativistic corrections to the one-body operators
together with the relativistic kinematics describing the motion of the
emitted nucleon. The dashed lines show the non relativistic FG
results.
}
\label{relfg}
\end{figure}

A prescription to consistently implement the expressions
(\ref{relcharge}) and  (\ref{relcurrent}) within a non relativistic
description of the nucleus has been presented in ref. \cite{Ama96a}. 
The basic idea consists in calculating the wave function of the
emitted particle of energy $\epsilon_p=\epsilon_h+\omega$ by inserting
in the Schr\"odinger equation the relativistic version of the kinetic
energy operator. This is equivalent to solving the Klein-Gordon equation
instead of the Schr\"odinger one. The asymptotic momentum of the
ejected nucleon is given by:
\begin{equation}
p^2= (\epsilon_p+M)^2-M^2 = \epsilon_p(\epsilon_p+2M) 
  = 2M\left[ \omega\left(1+\frac{\omega}{2M}\right)
            +\epsilon_h\left(1+\frac{2\omega+\epsilon_h}{2M}\right)
      \right]
\end{equation}
In the non-relativistic regime $\omega$ is much smaller than the
nucleon mass, therefore the $1/M$ terms can be neglected and the non 
relativistic relation $p^2=2M(\omega+ \epsilon_h)$ is recovered. A 
better approximation is obtained by neglecting only the 
$\epsilon_h \displaystyle \frac{2\omega+\epsilon_h}{2M}$ term.
This approximation can be implemented 
in a non relativistic model by making the substitution 
$\omega\rightarrow \omega(1+\omega/2M)$.
This modification of $\omega$ should be used only for the calculations
of the the responses \cite{Alb90}, but {\em not} in the nucleon form
factors, because the energy and momentum transfer are still $q$ and
$\omega$ (the only difference is that the momentum of the ejected
particle is calculated by using the relativistic energy-momentum
relation). 

It is possible to test the validity of this prescription by comparing
the results obtained with those of the relativistic FG model. In this
last model the sp wave functions are free solutions of the Dirac
equation and the relativistic expressions of the current operators can
be used without doing any approximation in the calculations of the
responses. We show in fig. \ref{relfg} this comparison and we see
that, even at high values of $q$ where the relativistic effects are more
important, the agreement between the results obtained with the
prescription presented above and those of the relativistic FG is
excellent. The prescription can also be used in CMS calculations and,
for high $q$ where surface effects are negligible, good agreement with
the relativistic FG model results has been found \cite{Ama96a}.

This procedure for including relativistic effects has been successfully
applied to the study (e,e'p) processes from polarized deuterium
\cite{Jes98} and heavier nuclei \cite{Ama96b,Ama98a}. More recently
the same expansion procedure has  been applied to the MEC, although in
this case the complexity of these current operators make the calculations more
involved. The first results \cite{Ama98c} indicate a simple way of
correcting the MEC for relativistic effects by using meanly
multiplicative factors which depend on $q$ and $\omega$. 

From the above investigation it turns out that the relativistic
kinematics produce a reduction of the width of the inclusive one-body
responses and a shift of the peak positions towards smaller values of
$\omega$. In fact, the peak position is located at
$\omega=(q^2-\omega^2)/2M$, which for $\omega \ll M$ can be
approximated by $\omega=q^2/2M$, a larger value. The reduction of the
width of the QE responses is noticeable in the upper limit of the
$\omega$ region already for $q=500$~MeV. For instance, for a value of
$k_{\rm F}=235$~MeV, about the average density of $^{40}$Ca the
maximum allowed value of $\omega$ is $\sim 225$~MeV in the
relativistic FG, against  $\sim 260$~MeV in the non relativistic FG
(see ref. \cite{Ama96a} for details). 

In summary, the relativistic corrections to the electromagnetic
nuclear responses are important for high $q$ and the model developed
for ``relativizing'' the currents can account for these corrections in
a way which is easy to implement in already existing non-relativistic
models. 
 
\subsection{Random Phase Approximation}
\label{rpa}
In the MF model the many-body hamiltonian is a sum of single particle
hamiltonians. This assumption neglects part of the interaction between
nucleons which, for this reason, is called residual interaction. Under
certain approximations, the effects of the residual interaction on the
description of the exited states of the many-body systems are
considered within the Random Phase Approximation (RPA).

The basic hypothesis of the RPA is that the nuclear excited state
$|f \rangle$ can be described as a linear combination of particle-hole
(p-h) and hole-particle (h-p) excitation of the ground state 
$|i \rangle$:
\begin{equation}
\label{rpastate} 
|f \rangle=\sum_{ph} \left( X^f_{ph} a^+_p a_h - Y^f_{ph}
 a^+_h a_p \right) |i \rangle \, ,
\end{equation}
where $a^+$ and $a$ are the creation and annihilation operators and 
$p$ ($h$) indicates a particle (hole) state. The RPA amplitudes $X$
and $Y$ are provided by the theory and are normalized such as:
\begin{equation}
\label{rpanorm} 
\sum_{ph} \left[ (X^f_{ph})^2  - (Y^f_{ph})^2 \right]=1 .
\end{equation}

Once the $X$ and $Y$ RPA amplitudes are known, the expression
(\ref{rpastate}) is used in the eqs. (\ref{respl}) and (\ref{respt})
to calculate the transition amplitudes and the responses. In analogy
to what happens in the shell model case, the evaluation of the
complicated many-body transition matrix element is reduced to the
calculation of a set of one-body transition matrix elements whose
contributions for a specific transition are weighted by $X$ and $Y$.

The RPA has been implemented in nuclear physics in various ways, but
for the application to the QE region it is necessary to consider 
excitations to the continuum. This means that in eq. (\ref{rpastate})
the sum on $p$ should be understood as a sum on the $l_pj_p$ and $t_p$
quantum numbers characterizing the particle state and an integral on
its energy $\epsilon_p$. 

The RPA can be applied also to an infinite system of fermions, and in
fact it is in such a system that the RPA was first formulated 
\cite{Boh53}. Whence the quantum numbers characterizing the sp
states are the momenta. Thus the two sums of eq.
(\ref{rpastate}) become three dimensional integrals on the momenta.

Following the approach of Section \ref{mfmodel} we shall discuss
separately the RPA results in finite and infinite systems.

\subsubsection{Finite systems}
\label{finite}
The various implementations of the RPA to study nuclear excitations in
the continuum, the Continuum RPA (CRPA), can be classified in two
different approaches.

In a first approach, the nucleon-nucleon effective interaction, used
first in a Har\-tree-Fock calculation to generate the set of sp wave
functions, is also used to perform the CRPA calculation. This approach,
called self-consistent, is usually based upon the zero-range Skyrme
interaction \cite{Sky56}. In QE \cite{Cav84,Cav90,Van95} calculations
various parametrizations of this interaction have been used. All of
them reproduce quite well the ground state properties of the doubly
closed shell nuclei, and give a reasonable description of the giant
resonances \cite{Vau72,War87,Sar93}.

The second approach is based upon the Landau-Migdal theory of finite
Fermi systems \cite{Mig57}. This theory assumes a good knowledge of
the ground state of the system and the  RPA equations describe density
fluctuations around the equilibrium. In the case of nuclei the ground
state is usually described by means of a Woods-Saxon potential. Within
this approach various kind of interactions have been used: from the
traditional Landau-Migdal interaction of zero-range type up to finite
range interactions \cite{Bri87} like the polarization potential
\cite{Co'88} and the G-matrix \cite{Bub91,Jes94}. 
%
%
\begin{figure}
\begin{center}                                                                
\leavevmode
\epsfysize = 500pt
\hspace*{-0.65cm}
\makebox[0cm]{\epsfbox{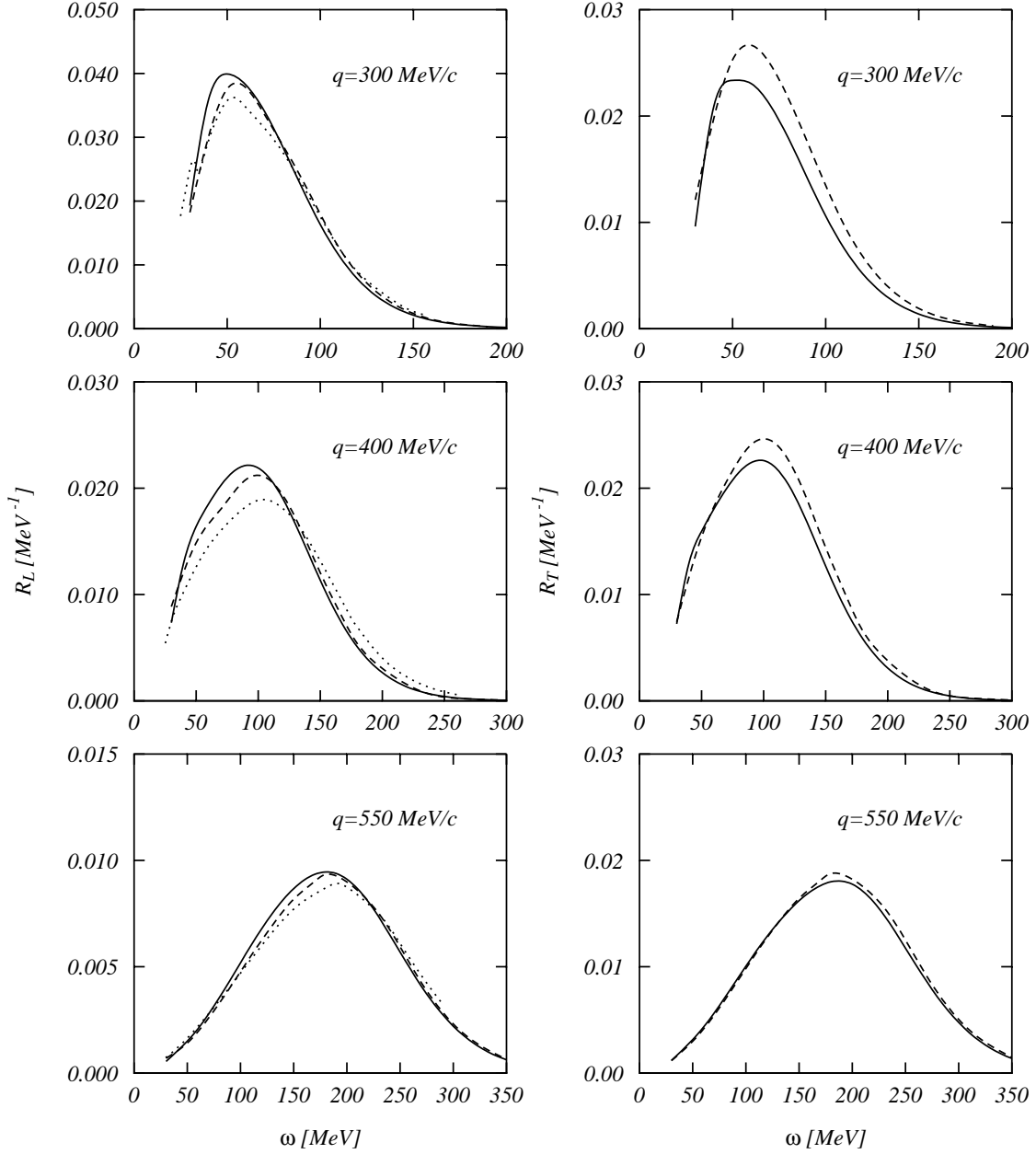}}
\end{center}
\vspace{-1.0 cm}
\caption{\small Responses calculated in CSM (full lines) and with
Continumm RPA. The dotted lines have been obtained with a zero range 
Landau-Midgal interaction \protect\cite{Co'84,Co'85} while the dashed
lines have been obtained using the finite range polarization potential
\protect\cite{Pin88a,Pin88b}.
}
\label{crpa}
\end{figure}

In fig. \ref{crpa} we show a comparison between MF longitudinal
responses (full lines) and CRPA responses calculated with the
Landau-Migdal interaction (dotted lines). The peaks of the CRPA
responses are  about 10\% smaller with respect to the peaks of the MF
responses. Furthermore their position is moved towards higher values
of the excitation energy. This effect is more pronounced at high $q$
values. Similar effects have been found also in self-consistent
calculations with the Skyrme interaction 
\cite{Cav84,Cav90,Van95}.

The applicability of zero-range interactions in the QE region is
quite doubtful because of the high values of $q$ and $\omega$. 
The zero-range interactions are constant in momentum space, while
finite-range interactions go to zero at high $q$. For this reason
calculations done with  zero-range interactions overestimate the RPA
effects in excited states at high $q$ and $\omega$. The implementation
of finite-range interactions in CRPA calculation is quite complicated
because of the need of including exchange diagrams which cannot be
easily handled in RPA. 

The polarization potential Pines et al. \cite{Pin88a,Pin88b} has the
nice feature of being a finite-range interaction constructed to be
used in RPA calculations with only direct terms. The effect of the
exchange terms is considered in an effective manner with an
appropriate choice of the parameters of the interaction. The dashed
lines of fig. \ref{crpa} show CRPA results obtained with the
polarization potential \cite{Co'88}. As expected the RPA effects are
smaller than those obtained with the zero-range interaction and they
become negligible with increasing values of $q$. More elaborated CRPA
calculations which consider  finite-range interactions in both direct
and exchange terms \cite{Del85,Bri87,Shi89,Bub91,Jes98} produce
analogous results.

In spite of the differences in the technology of the solution of the
CRPA equations and in the input, the various calculations in the QE
region show that the RPA correlations lower the longitudinal responses
and enhance the transverse ones (see right panels in fig. \ref{crpa}).
When a finite-range interaction is used, the RPA corrections to the
MF results are quite small, within the uncertainty band related to
the choice of the nucleon form factors.

\subsubsection{Infinite systems}
\label{infinite}

In infinite systems the QE excitations are often calculated
\cite{Alb84,Alb87a,Alb87b,Alb89,Alb90,DeP93,Gil97} using the
so-called Ring Approximation (RA). The RA is a RPA where the exchange
diagrams are neglected. This allows one to obtain very compact
expressions. In RA, the polarization propagator $\Pi$, whose imaginary
part is proportional to the response, can be calculated as:
\begin{equation}
\label{ring} 
\Pi^{\rm RA}(\nq,\omega)=
\frac{\Pi^0(\nq,\omega)}{1-V_{ph}(\nq,\omega)\Pi^0(\nq,\omega)},
\end{equation}
where $V_{ph}$ is the effective nucleon-nucleon interaction
and $\Pi^0$ the free propagator.

\begin{figure}[ht]
\vspace{-3.cm}
\begin{center}                                                                
\leavevmode
\epsfysize = 500pt
\makebox[0cm]{\epsfbox{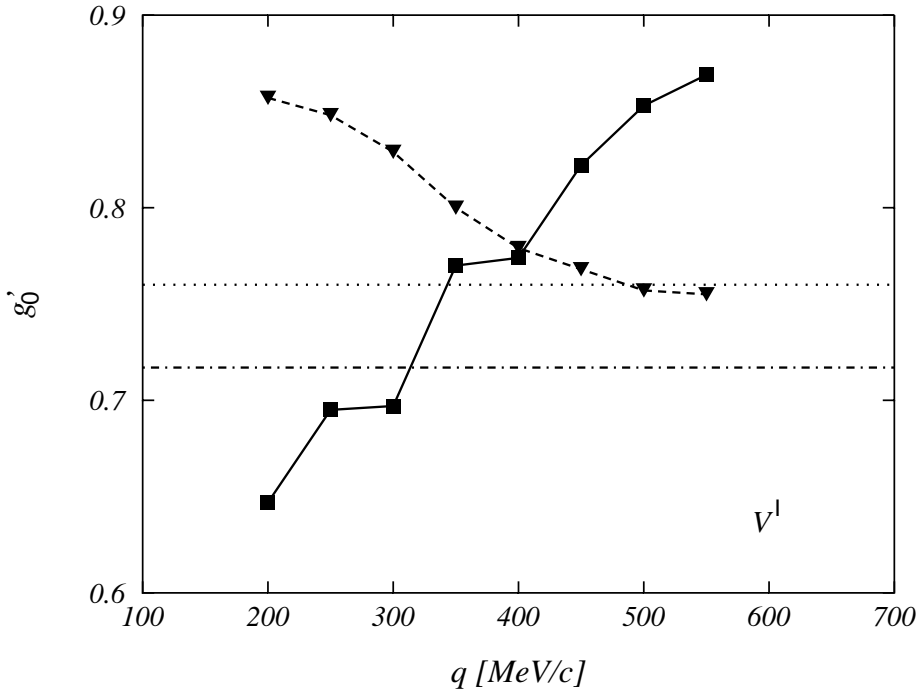}}
\end{center}
\vspace{-5.0 cm}
\caption{\small  Momentum dependence of the
values of the parameter $g_0^\prime$ to be used in RA calculations in
order to reproduce the peak positions (squares) and the non-energy
weighted sum rule (triangles) corresponding to the RPA responses. The
dotted line gives the value $g_0^\prime=0.76$ used in the RPA
calculation. The dashed-dotted line shows the value $g_0^\prime=0.717$
which reproduces some low-energy properties by means of a RA
calculation.
}
\label{ra}
\end{figure}

Since the RA calculations in an infinite system are rather easy, it
has been commonly accepted that RPA results can be simulated by RA by
properly changing the parameters of the effective interaction. This 
issue has been investigated in refs. \cite{Bau98a,Bau98b} where
RPA and RA transverse responses have been compared.
The RPA responses have been calculated using a residual interaction
containing a finite range term produced by the exchange of the $\pi$ and 
$\rho$ meson plus a zero-range term of Landau--Migdal type. The RA
calculations have been done using the same interaction but modifying 
the spin--isospin term of the zero range part, the $g_0^\prime$ parameter, 
in order to best approximate the RPA results. The comparison of the two
calculations has been done for various values of the momentum transfer
$q$. The values of $g_0^\prime$ needed to reproduce the peak positions of
the RPA responses (squares) and those needed to reproduce the non
energy weighted sum rule (triangles) are shown in fig. \ref{ra}. 
Only around $q=$400 MeV/c it is possible to find a value of $g_0^\prime$
able to reproduce both properties, but in general the results of the
two theories are incompatible. 

%
%

The choice of  the effective nucleon-nucleon interaction remains 
a major problem even in full RPA calculations for infinite
systems. 
In finite system RPA calculations the interaction is
fixed to reproduce properties of low-lying excited states and, within
the same RPA approach, it is afterwards used to calculate the QE
response. This procedure cannot be used in infinite systems and
therefore the interaction is taken from finite systems RPA
calculations. From the theoretical point of view this procedure is
inconsistent since the effective interaction is properly defined only
within the framework of the effective theory where it is used. Usually
the situation is even worse since the interaction taken from finite
systems RPA calculations is used to perform RA calculations in
infinite systems. A detailed study of these inconsistencies has been
done in refs. \cite{Bau98a,Bau98b}.

In spite of these theoretical inconsistencies the RA results  show
effects rather similar to those of the CRPA calculations. The reason
of this agreement is due to the small influence of the effect
investigated especially at high $q$ values.

\subsection{Final state interaction}
\label{fsisrt}
In any nuclear disintegration process the emitted nucleons interact
with the residual nucleus. This obvious physical effect, which takes
the name of Final State Interaction (FSI), is not considered in a FG
model where the wave function of the emitted nucleon is a plane wave. 
In the CSM the emitted particles interact with the rest nucleus
through the average nuclear potential. Strictly speaking this is part
of the FSI. The CRPA calculations consider the mixing of the various
nucleon emission channels and this is another part of the FSI. In the 
literature, the name FSI has been used to indicate those effects on
the final state of the system which are beyond the MF and RPA
descriptions. In this section we shall discuss them.
%
%
\begin{figure}
\vspace{-2.cm}
\begin{center}                                                                
\leavevmode
\epsfysize = 350pt
\makebox[0cm]{\epsfbox{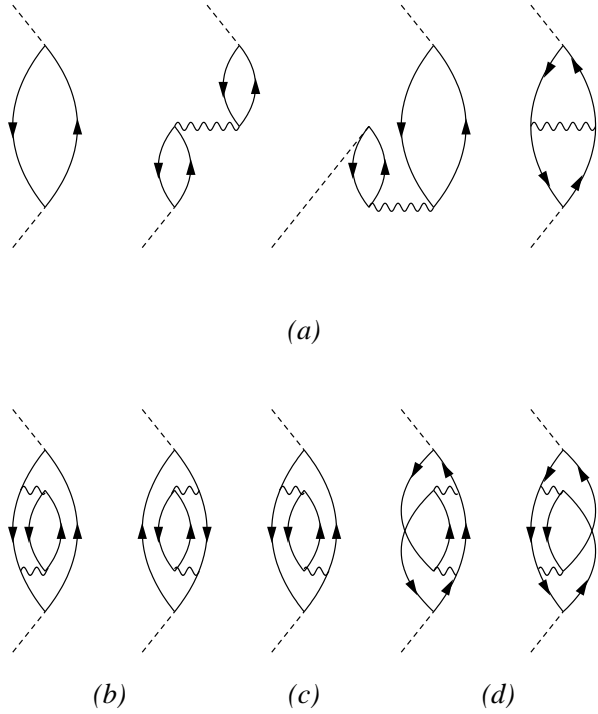}}
\end{center}
\vspace{-2.5cm}
\caption{\small Representative low-order diagrams iterated in the second RPA
theory. The (a) diagrams are the usual RPA diagrams. 
The (b),(c) and (d) diagrams are produced by the 2p-2h excitations. 
In addition to the self-energy insertions on a single particle line
(b), there are ph linked diagrams of bubble (c) and ladder (d) type.
}
\label{rpadiag}
\end{figure}
%
%
%
\begin{figure}
\vspace{-5.0cm}
\begin{center}                                                                
\leavevmode
\epsfysize = 500pt
\hspace*{1.5cm}
\makebox[0cm]{\epsfbox{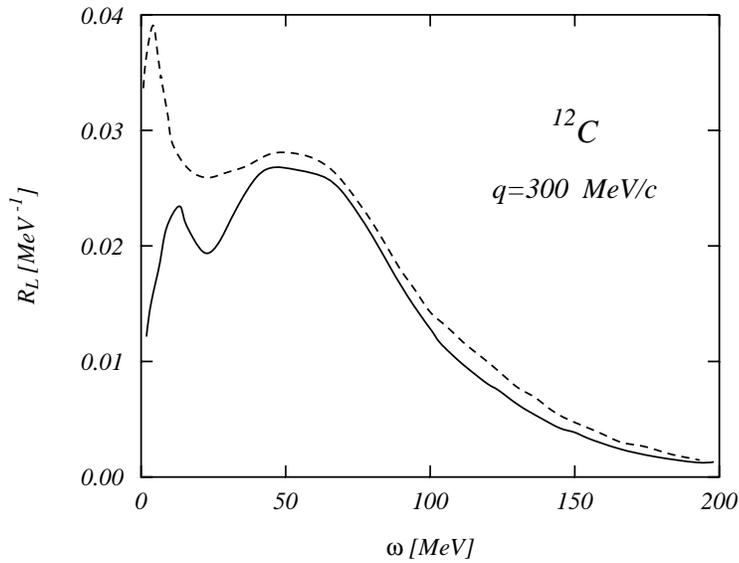}}
\end{center}
\vspace{-4.cm}
\caption{\small Longitudinal response at q=300~MeV/$c$ in $^{12}$C. 
The full line shows the full second RPA result. The dashed line has 
been obtained by neglecting the ph linked diagrams, the (c) and (d)
diagrams of fig. \ref{rpadiag}.
}
\label{srpa}
\end{figure}

A microscopic description of the FSI should be done with theories
beyond the RPA. One of these theories is the second RPA
\cite{Yan82,Yan83,Dro90} which describes the nuclear excited state as
a linear combination of 1p-1h and 2p-2h excitations. In
ref. \cite{Dro87} second RPA calculations in the QE region have been
performed in the $^{12}$C nucleus for low values of the momentum
transfer. In these calculations the continuum has been
discretized. The excessive
numerical effort needed imposed a limitation of the
configuration space used, which turned out to be too small to provide a
realistic description of the QE excitation. In spite of this
limitation, the results obtained produced important information from
the theoretical point of view.

In fig. \ref{rpadiag} the low-order diagrams iterated in the second RPA
calculation are shown. The diagrams (a) are the usual RPA diagrams,
while the other ones are present  only in a second RPA
calculation. The (b) diagrams correspond to a correction of the sp
wave function, both hole and particle, produced by a self-energy
insertion. The effects produced by these diagrams can be simulated by
inserting an imaginary part in the MF potential.

In fig. \ref{srpa} we present the $^{12}$C longitudinal response at 
q=300~MeV/$c$ obtained with a complete second RPA calculation (full
line) and by neglecting the (c) and (d) diagrams of fig. \ref{rpadiag} 
(dashed line). The similarity of the two curves in the QE region
indicates that the most important diagrams beyond RPA are those of the
self-energy insertion. This result is not valid in the giant resonance
region \cite{Dro90} as it is shown  by the low energy part of
fig. \ref{srpa}. In ref. \cite{Dro89} the effect of the (c) and (d)
diagrams has also been investigated on the transverse response and it
has been found to be slightly greater than in the longitudinal
response.

In spite of this difference the results of refs. \cite{Dro87,Dro89}
indicate that the second RPA effects in the QE region can be simulated
reasonably well by a complex optical potential. This approach has
been used in MF \cite{Hor80,Chi89,Cap91} as well as in RPA
\cite{Del85,Bri87,Bou89,Sag89,Bou91,Jes94} calculations. The imaginary
part of the optical potential removes part of the flux of the emitted
particles and therefore this reduces the responses. The use of a
complex, and energy dependent, optical potential has the disadvantage
that the sp wave functions are no longer orthogonal, thus they
do not form an orthonormal basis and cannot be used 
in many-body calculations like
RPA. Furthermore the sum rules, controlling the total excitation
strength, are no longer satisfied. These problems are usually
neglected under the assumption that non-orthogonality effects are
small. 
 
%
%
\begin{figure}
\begin{center}                                                                
\leavevmode
\epsfysize = 500pt
\hspace*{-.65cm}
\makebox[0cm]{\epsfbox{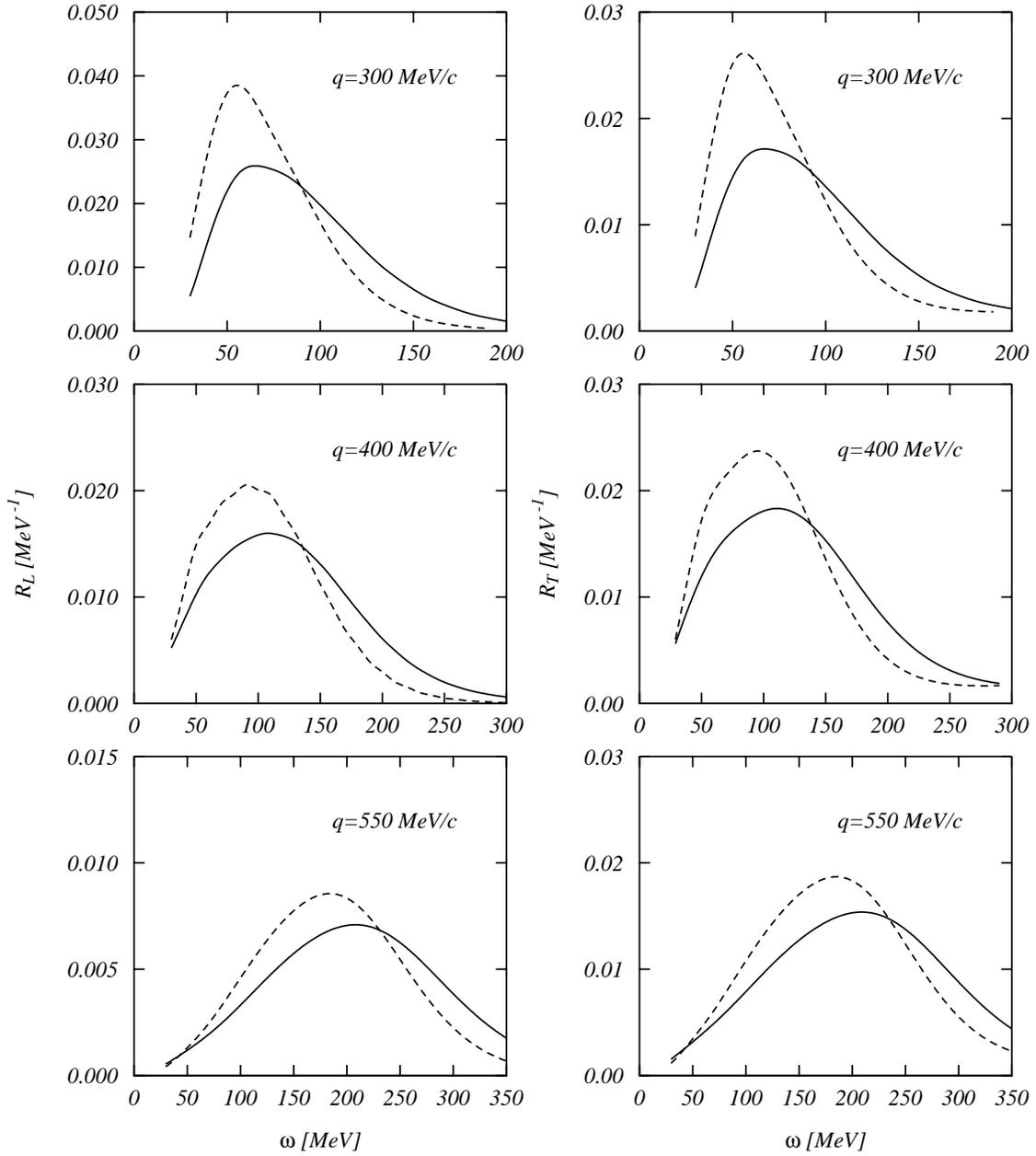}}
\end{center}
\vspace{-1.0 cm}
\caption{\small $^{12}$C responses obtained with CSM (dashed lines). 
The full lines have been obtained applying the FSI corrections.
}
\label{smooth}
\end{figure}

A different procedure to treat the FSI has been proposed in
ref. \cite{Co'88}. The basic assumption is to consider the self-energy
of the system independent from the sp states: $\Sigma(\omega)=
\Delta(\omega) + i \Gamma(\omega)/2$. The response containing FSI
effects, $R^{FSI}$, can be obtained by a folding procedure: 
\begin{equation}
\label{fsiresp} 
R^{\rm FSI}(q,\omega)= \int_0^\infty \, {\rm d}E \, R(q,E) \,
\left[ \rho(E,\omega) + \rho(E,-\omega) \right]
\end{equation}
where the function $\rho$ is related to the real and imaginary parts
self-energy as follows:
\begin{equation}
\label{fsirho} 
\rho(E,\omega)= \frac{1}{2\pi} \,
\frac{\Gamma(\omega)}
{\left[ E-\omega -\Delta(\omega)\right]^2 + 
\left[\Gamma(\omega)/2\right]^2}.
\end{equation}
The $\Gamma$ and $\Delta$ functions are fixed to reproduce the
empirical values of the volume integrals of the optical potential
\cite{Mah82}. In this treatment of the FSI the sum rules are conserved
since the strength of the response is not reduced but it is
redistributed on the energy spectrum. The general effect of FSI is a
widening of the response width and a lowering of the peak value.

In ref. \cite{Co'88} a $q$ dependent nucleon effective mass $m^*$ such
that $m^*/m<1$, has been introduced to consider the non-locality of the
static mean field. For a given $m^*/m$ value the following scaling
relation holds:
\begin{equation}
R_{m^*}(q,\omega)= \frac{m^*}{m}\, 
R\left( q, \frac{m^*}{m} \omega \right) \, ,
\end{equation}
which shows that the response is lowered and that its maximum is
shifted to higher energy.

The inclusion of the effective mass still conserves the non-energy
weighted sum rule, but it violates the energy weighted one. 
The size of this violation can be quantified by a relative enhancement
factor which is of about 0.2 at $q$=300~MeV/$c$ and 0.13 at $q$=550
MeV/$c$ \cite{Co'88}. 

The total effect of FSI and effective mass on the CSM responses of
 $^{12}$C is shown in fig. \ref{smooth}. These results are rather
similar to those obtained using the optical potential.

It is worth remarking that among the various corrections to the MF
responses the FSI produces the biggest ones. It is curious, and
unexpected, that effects related to the many-particles many-hole
excitations of the nucleus are more important than those related to
1p-1h excitations (like RPA). This fact can be understood by
considering the strength of the residual interaction and the density
of nuclear final states. The residual interaction in the QE region is
very weak, therefore RPA effects are small. On the other hand, the
great number of 2p-2h states available in this region compensates the
weakness of the residual interaction and this produces sizeable
effects on the response.

\section {Comparison with experimental data}
\label{experiment}
Electron scattering experiments in the QE region have been performed
at different laboratories for various nuclear targets. Longitudinal
and transverse responses have been separated in $^2$H \cite{Ber87},
$^3$H \cite{Ber87,Dow88}, $^4$He \cite{Ber87,Zgh94}, $^{12}$C
\cite{Bar83}, $^{40}$Ca \cite{Dea83,Mez84,Mez85,Dea86,Yat93,Wil97},
$^{48}$Ca \cite{Mez84,Mez85,Dea86}, $^{56}$Fe
\cite{Alt80,Mez84,Hot84,Mez85}, $^{208}$Pb \cite{Zgh94} and $^{238}$U
\cite{Bla86}.

\begin{figure}
\begin{center}                                                                
\leavevmode
\epsfysize = 500pt
\hspace*{-.65cm}
\makebox[0cm]{\epsfbox{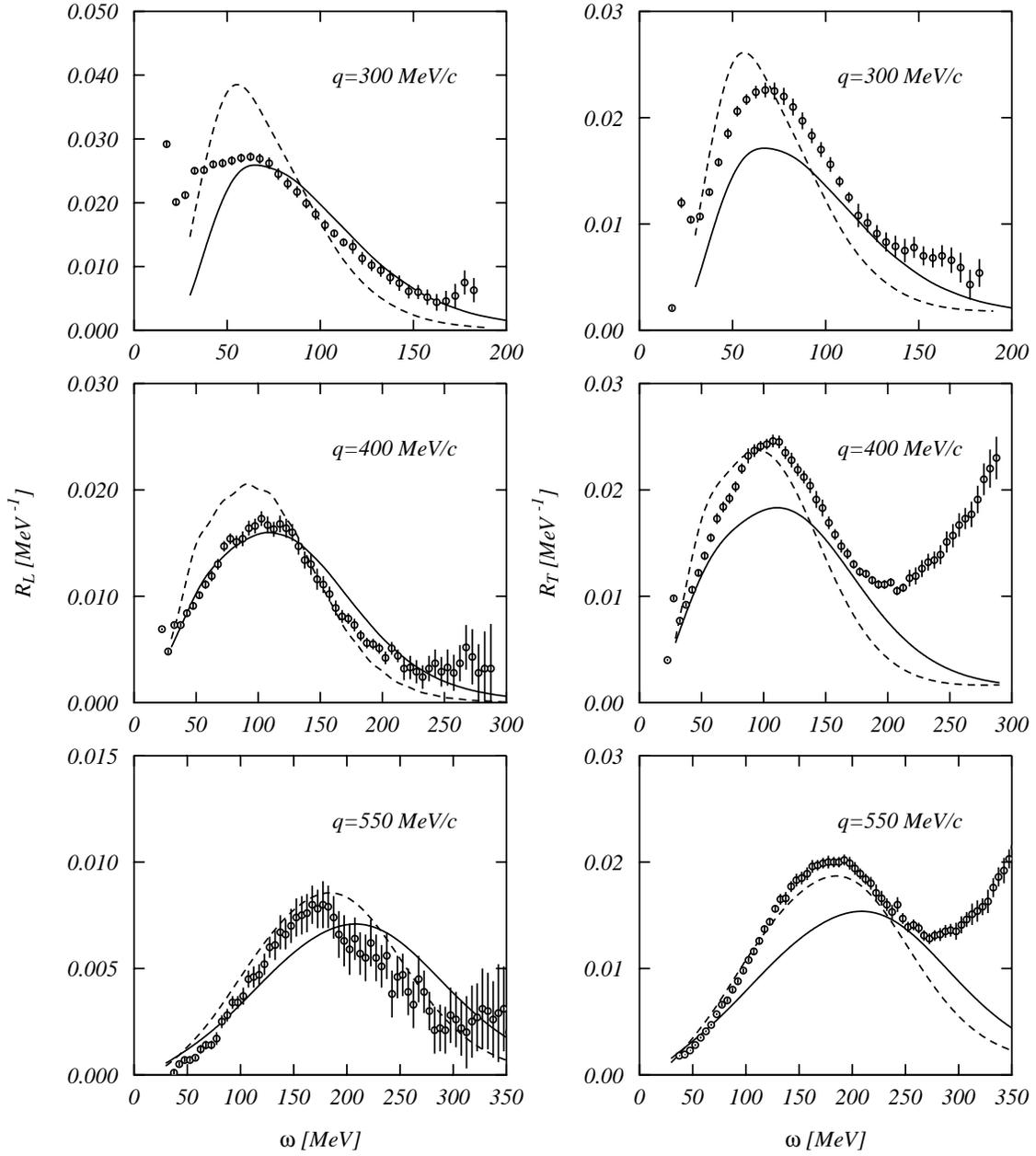}}
\end{center}
\vspace{-1cm}
\caption{\small
Comparison of the  $^{12}$C
experimental data of ref. \protect\cite{Bar83} with
the CSM responses (dashed lines) and those 
obtained including the various corrections discussed in this paper,
excluding the relativistic corrections.
}
\label{c12exp}
\end{figure}

\begin{figure}
\begin{center}                                                                
\leavevmode
\epsfysize = 500pt
\hspace*{-.65cm}
\makebox[0cm]{\epsfbox{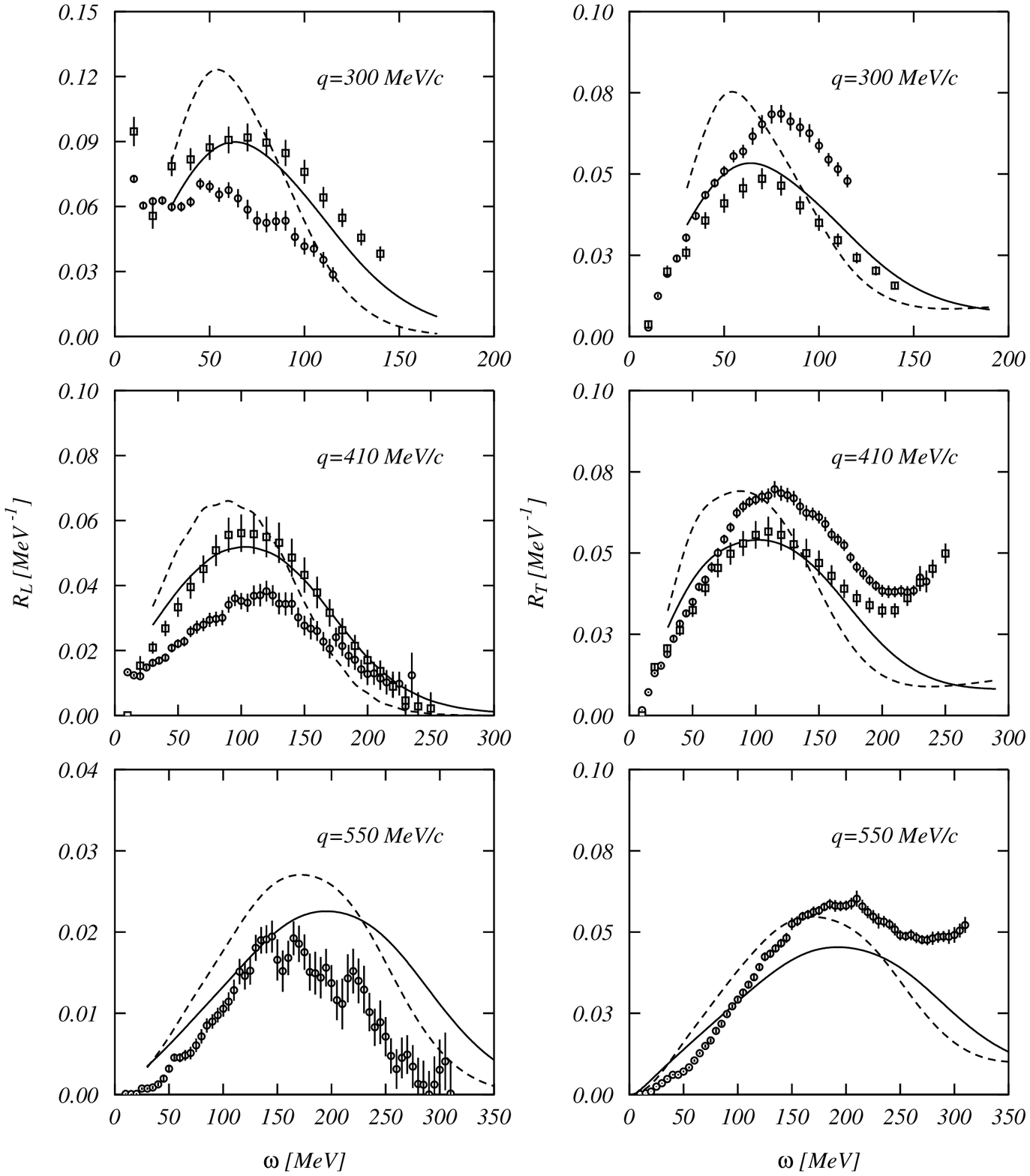}}
\end{center}
\vspace{-1cm}
\caption{\small The same as the previous figure for $^{40}$Ca. The
data are from refs. \protect\cite{Mez84,Mez85} (circles) and
\protect\cite{Yat93,Wil97} (squared). 
}
\label{ca40exp}
\end{figure}

The comparison between theory and experiments is satisfactory in
few-body systems, but for medium-heavy nuclei it is not possible to
reproduce at the same time both responses. As example we show in
figs. \ref{c12exp} and \ref{ca40exp} the comparison between the
results of our calculations with the available data on $^{12}$C and
$^{40}$Ca. Each panel shows two curves, a dashed one corresponding to
the standard treatment, i.e. CSM calculation with one body currents
only, and a continuum curve containing all the corrections discussed
in this paper excluding the relativistic effects.

%
%
%
%
%
%
%

In the standard treatment the longitudinal responses are overestimated
while the transverse ones are reasonably well reproduced. This fact
provoked disconcern since the electromagnetic operator generating the
longitudinal response, the charge operator, is supposed to be better
understood than the current operator.

A great theoretical effort has been done to explain the reduction of
the longitudinal response. Various possibilities have been
investigated, from relativistic effects
\cite{Cel85,Bro89a,Weh89,Hor90} to  modifications of nucleon
properties in the medium \cite{Nob81,Nob83,Bro89b}. It is now
generally accepted that the FSI is the mechanism responsible for the
lowering of the longitudinal response. The same effect acts also
on the transverse response destroying the previous agreement (as is
shown by the full lines of the figs. \ref{c12exp} and
\ref{ca40exp}. The problem now has been moved from the longitudinal to
the transverse response whose empirical values are larger than those
predicted.

A word of caution is necessary at this point. The responses are not
directly measured but they are extracted from a set of cross section
measurements. The extraction procedure is rather complicated and it is
based upon various theoretical hypotheses on the structure and the
behavior of the cross section. Some of the published data have been
modified after a reanalysis \cite{Jou95,Jou96}. The discussion on the
reliability of the existing data and on the need of acquiring new ones
is still open. It is not our intention to discuss this issue here.  

We would like to remark however that the new set of $^{40}$Ca data
\cite{Yat93,Wil97} (squares in fig. \ref{ca40exp}) agrees rather well
with our calculations.

\section {Conclusions}
\label{conclusions}
In this article we have discussed the validity of the approximations
made in what we have called the standard description of the QE excitation.
This is a PWBA calculation done within non relativistic MF nuclear
model and using  one-body currents only.

The PWBA is usually adopted because it leads to a simple
expression of the cross section, which allows the separation of
longitudinal and transverse responses through a Rosenbluth
procedure. Distortion effects produced by the Coulomb field of the nucleus
are not very big. For nuclei with $Z<20$ they can be rather well
simulated by the effective momentum transfer prescription
(\ref{effmom}). For heavier nuclei the focusing of the electron wave
function onto the nucleus becomes relevant and DWBA calculations are
necessary \cite{Co'87a,Tra88}.

A major source of uncertainty in the electron-nucleus interaction is
related to the choice of the electromagnetic nucleon form factors
\cite{Ama93a}. We have shown in fig. \ref{form} the uncertainty band
produced by the various possible choices.

The size of the two-body currents effects is within this uncertainty 
band. We have studied two-body currents produced by the exchange of
one pion. We have seen that the total effect of these MEC on the
response is small because of the cancelation between the seagull and
pionic current contributions \cite{Ama93b}. Seagull and pionic
currents modify the response in opposite directions and the final
effect is rather small. We have also evaluated the contribution of the
currents associated to the excitation of a virtual $\Delta$ which has
been found  to lower the transverse response
\cite{Ama94a} but to be rather small.

An estimate of the effects produced by relativistic corrections shows
that their size is within the form factor uncertainty band as long as
the momentum transfer value remains smaller than 500~MeV/$c$
\cite{Ama96a,Ama96b,Ama98a}.
 
We started our investigation of the nuclear excitation mechanism by
testing the need of using a finite system description for the QE
excitation. We have compared CSM and FG responses and we found that
they are rather similar if the value of the Fermi momentum is
evaluated by considering the average nuclear density (see eq. 
(\ref{kav})). The agreement worsen by using the LDA \cite{Ama94b}.

We have seen that RPA corrections are not important when a
finite-range interaction is used \cite{Co'88,Ama93b}. Zero-range
interactions overestimate RPA effects, especially at high values of
the momentum transfer. In infinite nuclear systems the choice of the
residual interaction is inconsistent with the effective theory where
it is used \cite{Bau98a,Bau98b}. From a pragmatic point of view these
uncertainties are irrelevant since RPA effects are small.

The largest correction to the standard description of the QE
excitation is produced by the FSI. Microscopic calculations done
within the second RPA framework indicates that, in the QE region, the FSI
can be rather well described in terms of an imaginary part of the optical
potential \cite{Dro87}. We have developed an alternative approach to
deal with FSI \cite{Co'88}. Our results are similar to those obtained
using the optical potential. The FSI enlarges the widths of the
responses and lowers the peak values \cite{Ama93b}.

The comparison with the experimental $^{12}$C data \cite{Bar83} shows
that, when the FSI corrections are included, the longitudinal
responses are reasonably well reproduced while the transverse one are
underestimated \cite{Ama94a}. The same kind of calculation in
$^{40}$Ca does not reproduce the Saclay data \cite{Mez84,Mez85} while
the new MIT data \cite{Yat93,Wil97} are rather well described in both
longitudinal and transverse responses.

There are various problems still unsolved in the study of the QE
excitation. At present the longitudinal responses seem rather well
understood but the transverse ones are underestimates. There are
exclusive data \cite{Ulm87,Ber90} indicating that the two-nucleon
emission in the QE region could be larger than that predicted by the
MEC \cite{Ama93b}. This suggest the need of considering short-range
correlations in the nuclear model
\cite{Fan87,Fab89,Lei90,Orl91,Fab94,Fab97,Ama98b,Co'98}. Also the
so-called dip region, the region between the QE peak and the peak of
the $\Delta$ resonance, is at present not well described.

The QE excitation of nuclei is a good testing ground of our knowledge
of the electron-nucleus interaction and our understanding of the
nuclear excitation mechanism. Since surface effects are not important
\cite{Ama94b} it is meaningful to compare directly the experimental
data with the results of complicated many-body theories developed in
infinite systems \cite{Fan87,Alb89,Fab89,Cen94,Fab94,Cen97,Fab97}.

We believe that the solution of the problems mentioned above and the
interest of the theoreticians in comparing the results of their models
with QE data require new accurate and reliable data which could be
rather straightforwardly obtained at the new electron accelerator
facilities.

\section* {Acknowledgments}

This article summarizes a work which has been carried on for more than
ten years. It is a pleasure to thank all the people with whom we have
collaborated during this period:
M.B. Barbaro,
E. Bauer,
J.A. Caballero,
T.W. Donnelly,
S. Dro\.zd\.z,
A. Fabrocini,
E.M.V. Fasanelli,
J. Heisenberg,
S. Jeschonnek,
S. Krewald,
A. Molinari,
E. Moya de Guerra,
K.Q. Quader,
R. Smith,
J. Speth,
A. Szczurek
J.M. Ud\'{\i}as,
and J. Wambach.
A special thank to P. Rotelli for the critical reading of the manuscript.


\begin{thebibliography}{1234567}
\addcontentsline{toc}{chapter}{Referencias}

\bibitem[Alb84]{Alb84} W. Alberico, M. Ericson and A. Molinari, 
                       {\sl Ann. Phys. (N.Y.)} {\bf 154} (1984), 356.
\bibitem[Alb87a]{Alb87a} W.M. Alberico, P. Czerski, M. Ericson and
                        A. Molinari,
                       {\sl Nucl. Phys. A} {\bf 462} (1987), 269.
\bibitem[Alb87b]{Alb87b} W.M. Alberico, G. Chanfray, M. Ericson and
                        A. Molinari,
                       {\sl Nucl. Phys. A} {\bf 475} (1987), 233.

\bibitem[Alb89]{Alb89} W.M. Alberico, R. Cenni and A. Molinari,
                       {\sl Prog. Part. Nucl. Phys.} {\bf 23} (1989), 171.
\bibitem[Alb90]{Alb90} W.M. Alberico, T.W. Donnelly and A. Molinari,
                       {\sl Nucl. Phys. A} {\bf 512} (1990), 541.

\bibitem[Alt80]{Alt80} R. Altemus et al., {\sl Phys. Rev. Lett.}
                       {\bf 44} (1980), 965.
\bibitem[Ama92]{Ama92} J.E. Amaro, G. Co', E.M.V. Fasanelli
                         and A.M. Lallena,
                        {\sl Phys. Lett. B} {\bf 277} (1992), 249.
\bibitem[Ama93a]{Ama93a} J.E. Amaro, 
                        ``Efecto de los grados subnucleares de
                          libertad en la respuesta cuasiel\'astica
                          nuclear'', 
                         {\sl Ph.D. Thesis },
                          Universidad de Granada, 1993.
\bibitem[Ama93b]{Ama93b} J.E. Amaro, G. Co', and A.M. Lallena,
                       {\sl  Ann. Phys. (N.Y.)} {\bf 221} (1993), 306.
\bibitem[Ama94a]{Ama94a}  J.E. Amaro, G. Co', and A.M. Lallena,
                       {\sl  Nucl. Phys. A} {\bf 578} (1994), 365.
\bibitem[Ama94b]{Ama94b}  J.E. Amaro, A.M. Lallena and G. Co'
                       {\sl Int. Jour. Mod. Phys. E} {\bf 3} (1994), 735.
\bibitem[Ama96a]{Ama96a} J.E. Amaro, J.A. Caballero, T.W. Donnelly,
                         A.M. Lallena, E. Moya de Guerra and J.M. Ud\'{\i}as,
                       {\sl Nucl. Phys. A} {\bf 602} (1996), 263.
\bibitem[Ama96b]{Ama96b} J.E. Amaro, J.A. Caballero, T.W. Donnelly
                         and E. Moya de Guerra,
                       {\sl Nucl. Phys. A} {\bf 611} (1996), 163.
\bibitem[Ama98a]{Ama98a} J.E. Amaro and T.W. Donnelly,
                       {\sl Ann. Phys. (NY)} {\bf 263} (1998), 56.
\bibitem[Ama98b]{Ama98b} J.E. Amaro, A.M. Lallena, G. Co' and A. Fabrocini
                       {\sl  Phys. Rev. C} {\bf 57} (1998) 3473.
\bibitem[Ama98c]{Ama98c} J.E. Amaro, M.B. Barbaro, J.A. Caballero,
                         T.W. Donnelly and A. Molinari, 
                        {\sl Nucl. Phys. A} {\bf 643} (1998) 349

\bibitem[Bar83]{Bar83} P. Barreau et al., {\sl Nucl. Phys. A}
                        {\bf 402} (1983), 515;
                         {\sl Note} CEA-N-2334, Saclay (1983).
\bibitem[Bau98a]{Bau98a} E. Bauer and A.M. Lallena,
                         {\sl Phys. Rev. C} {\bf 57} (1998) 1681.
\bibitem[Bau98b]{Bau98b} E. Bauer and A.M. Lallena,
                         {\sl submitted to Phys. Rev. C}
\bibitem[Ber72]{Ber72} W. Bertozzi, J. Friar, J. Heisenberg and
                       J.W. Negele, {\sl Phys. Lett. B} {\bf 41} (1972),
                       408.
\bibitem[Ber87]{Ber87} A.M. Bernstein, {\sl Proceedings of the
                       3rd. Workshop on Perspectives in Nuclear Physics at 
                       Intermediate  Energies } (S. Boffi, C. Ciofi degli 
                       Atti and M. Giannini, Eds.), World
                       Scientific, Singapore, 1987.
\bibitem[Ber90]{Ber90} W. Bertozzi , {\sl Proceedings of the
                        Workshop on Two Nucleon Emission Reaction} 
                       (O. Benhar and A. Fabrocini, Eds.), ETS Editrice,
                        Pisa, 1990. , p.25
\bibitem[Bjo64]{Bjo64} J.D. Bjorken and S.D. Drell,
                       ``Relativistic Quantum Mechanics", 
                       McGraw-Hill, New York 1964.
\bibitem[Bla86]{Bla86} C. Blatchey et al., {\sl Phys. Rev. C} {\bf 34}
                       (1986), 1243.
\bibitem[Bof96]{Bof96} S. Boffi, C. Giusti, F.D. Pacati and M. Radici,
                       {\sl Electromagnetic Response of Atomic Nuclei}
                       ,Clarendon Press, Oxford, 1996 
\bibitem[Boh53]{Boh53} D. Bohm and D. Pines 
                       {\sl Phys. Rev}{\bf 92} (1953) 609.
\bibitem[Bou89]{Bou89} P.M. Boucher, B. Castel, Y. Okuhara and H. Sagawa,
                       {\sl  Ann. Phys. (N.Y.)} {\bf 196}
                       (1989), 150.
\bibitem[Bou91]{Bou91} P.M. Boucher and J.W. Van Orden,
                       {\sl  Phys. Rev. C} {\bf 43} (1991), 582.
\bibitem[Bri87]{Bri87} F.A. Brieva and A. Dellafiore, 
                       {\sl Phys. Rev. C} {\bf 36} (1987), 899.
\bibitem[Bro89a]{Bro89a} R. Brokmann, D. Drechsel, J. Frank and P.G. Reinhard, 
                         {\sl  Z. Phys. A} {\bf 332} (1989), 51.
\bibitem[Bro89b]{Bro89b} G.E. Brown and M. Rho, {\sl Phys. Lett. B}
                         {\bf 222} (1989), 324.
\bibitem[Bub91]{Bub91} M. Buballa, S. Dro\.zd\.z, S. Krewald and J. Speth,
                       {\sl Ann. Phys. (N.Y.)} {\bf 208} (1991), 346.
\bibitem[Cap91]{Cap91} F. Capuzzi, C. Giusti and F.D. Pacati,
                       {\sl Nucl. Phys. A} {\bf 524} (1991), 681.
\bibitem[Cav84]{Cav84} M. Cavinato et al., {\sl Nucl. Phys. A} 
                      {\bf 423} (1984), 376.
\bibitem[Cav90]{Cav90} M. Cavinato, M. Marangoni and A.M. Saruis, 
                       {\sl Phys. Lett. B} {\bf 235} (1990), 346.
\bibitem[Cel85]{Cel85} L.S. Celenza, A. Harindranath and C.M. Shakin, 
                     {\sl  Phys. Rev. C} {\bf 32} (1985), 248.
\bibitem[Cen94]{Cen94} R. Cenni and P. Saracco, 
                     {\sl Phys. Rev. C} {\bf 50} (1994) 1851. 
\bibitem[Cen97]{Cen97} R. Cenni, F. Conte and P. Saracco, 
                     {\sl Nucl. Phys. A} {\bf 623} (1997) 391. 

\bibitem[Che71]{Che71} M. Chemtob and M. Rho, {\sl Nucl. Phys. A}
                       {\bf 163} (1971), 1.
\bibitem[Chi89]{Chi89} C.R. Chinn, A. Picklesimer and J.W. Van Orden, 
                      {\sl  Phys. Rev. C} {\bf 40} (1989), 790; 1159.
\bibitem[Cio80]{Cio80} C. Ciofi degli Atti, {\sl Prog. Part. Nucl.
                       Phys.} {\bf 3} (1980), 163.
\bibitem[Co'84]{Co'84} G. Co' and S. Krewald, {\sl Phys. Lett. B}
                              {\bf 137} (1984), 145.
\bibitem[Co'85]{Co'85} G. Co' and S. Krewald, {\sl Nucl. Phys A} 
                        {\bf 433} (1985), 392. 
\bibitem[Co'87a]{Co'87a} G. Co'  and J. Heisenberg,
                       {\sl Phys. Lett. B} {\bf 197} (1987), 489.
\bibitem[Co'87b]{Co'87b} G. Co', A.M. Lallena  and T.W. Donnelly,
                       {\sl  Nucl. Phys. A} {\bf 469} (1987), 684.
\bibitem[Co'88]{Co'88} G. Co', K.Q. Quader, R. Smith and J. Wambach, 
                        {\sl Nucl. Phys. A} {\bf 485} (1988), 61.
\bibitem[Co'98]{Co'98} G. Co' and A.M. Lallena, 
                       {\sl Phys. Rev. C} {\bf 57} (1998) 145. 
\bibitem[Czy63]{Czy63} W. Czy\.z, {\sl Phys. Rev.} {\bf 131} (1963), 2141. 
\bibitem[Dea83]{Dea83} M. Deady et al., {\sl Phys. Rev. C} {\bf 28} (1983), 
                       631.
\bibitem[Dea86]{Dea86} M. Deady et al., {\sl Phys. Rev. C} {\bf 33} 
                      (1986), 1897.
\bibitem[deF66]{deF66} T. deForest and J.D. Walecka, 
                       {\sl Adv. Phys.}  {\bf15} (1966), 57.
\bibitem[DeJ87]{DeJ87} C.W. DeJager and C. DeVries, {\sl At. Data and
                       Nucl. Data Tables} {\bf 36} (1987), 495.
\bibitem[Del85]{Del85} A. Dellafiore, F. Lenz and F.A. Brieva,
                       {\sl Phys. Rev. C} {\bf 31} (1985), 1088.
\bibitem[DeP93]{DeP93} A. DePace and M. Viviani,
                       {\sl Phys. Rev. C} {\bf 48} (1993) 2931.
\bibitem[Don75]{Don75} T.W. Donnelly and J.D. Walecka,
                       {\sl Ann. Rev. Nucl. Science} {\bf 25} (1975), 329.
\bibitem[Dow88]{Dow88} K. Dow et al., {\sl Phys. Rev. Lett.}
                       {\bf 61} (1988), 1706.
\bibitem[Dre89]{Dre89} D. Drechsel and M.M. Giannini,
                       {\sl Rep. Prog. Phys.} {\bf 52} (1989) 1083.
\bibitem[Dro87]{Dro87} S. Dro\.zd\.z, G. Co', J. Wambach and J. Speth,
                      {\sl Phys. Lett. B} {\bf 185} (1987), 287.
\bibitem[Dro89]{Dro89} S. Dro\.zd\.z, M. Buballa, S. Krewald and J. Speth,
                       {\sl Nucl. Phys. A} {\bf 501} (1989), 487.
\bibitem[Dro90]{Dro90} S. Dro\.zd\.z, S. Nishizaki, J. Speth and J. Wambach
                       {\sl Phys. Rep.}{\bf 197} (1990) 1.
\bibitem[Edm57]{Edm57} A.R. Edmonds,``Angular momentum in quantum
                       mechanics'', Princeton University Press, 
                       Princeton, 1957.
\bibitem[Fan87]{Fan87} S. Fantoni and V.R. Pandharipande, 
                       {\sl Nucl. Phys. A} {\bf 473} (1987), 234.
\bibitem[Fab89]{Fab89} A. Fabrocini and S. Fantoni,
                       {\sl  Nucl. Phys. A} {\bf 503} (1989), 375.
\bibitem[Fab94]{Fab94} A. Fabrocini,
                       {\sl Phys. Lett. B} {\bf 322} (1994) 171. 
\bibitem[Fab97]{Fab97} A. Fabrocini,
                       {\sl Phys. Rev. C} {\bf 55} (1997) 338. 
\bibitem[Fri77]{Fri77} J.L.Friar, {\sl Phys. Lett. B} {\bf 69} (1977), 51.
\bibitem[Gar76]{Gar76} M. Gari and H. Hyuga, {\sl Z. Phys. A} {\bf 277}
                         (1976), 291; {\sl Nucl. Phys. A}  {\bf 274}
                         (1976), 333.
\bibitem[Gar92]{Gar92} C. Garcia-Recio, J. Navarro, N. Van Giai and 
                       L.L. Salcedo,
                       {\sl Ann. Phys. (NY)} {\bf 214} (1992) 293. 
\bibitem[Gil97]{Gil97} A. Gil, J. Nieves and E. Oset,
                       {\sl Nucl. Phys. A} {\bf 627} (1997) 543.
\bibitem[Hei83]{Hei83} J. Heisenberg and H.P. Blok,
                       {\sl Ann. Rev. Nucl. Part. Sc.} {\bf 33} (1983),
                       569
\bibitem[Hoc73]{Hoc73} J. Hockert, D.O. Riska, M. Gari and
                        A. Huffman, {\sl Nucl.Phys. A} {\bf 217} (1973), 14.
\bibitem[Hoe76]{Hoe76} G.  Hoehler  et al., {\sl Nucl. Phys. B} 
                        {\bf 114} (1976), 505.
\bibitem[Hor80]{Hor80} Y. Horikawa, F. Lenz and N.C. Mukhopadhyay,
                       {\sl Phys. Rev. C}{\bf 22} (1980) 461. 
\bibitem[Hor90]{Hor90} C.J. Horowitz and J. Piekarewicz, 
                        {\sl Nucl. Phys. A} {\bf 511} (1990), 461.
\bibitem[Hot84]{Hot84} A. Hotta et al., {\sl Phys. Rev. C} {\bf 30} (1984), 87.


\bibitem[Iac73]{Iac73} F. Iachello, A.D. Jackson and A. Lande,
                       {\sl Phys. Lett. B}  {\bf 43} (1973), 191.

\bibitem[Jan66]{Jan66} T. Janssens, R. Hofstadter, E.B. Hughes
                       and M.R. Yearian, {\sl Phys. Rev.} {\bf 142}
                       (1966), 922.
\bibitem[Jes94]{Jes94}  S. Jeschonnek, A. Szczurek, G. Co' and
                        S. Krewald,
                        {\sl Nucl. Phys. A} {\bf 570} (1994) 599.
\bibitem[Jes98]{Jes98}  S. Jeschonnek and  T. W. Donnelly,
                        {\sl Phys. Rev. C}{\bf C57} (1998) 2438.
\bibitem[Jou95]{Jou95} J. Jourdan, {\sl Phys. Lett. B} {\bf 353} (1995) 189.
\bibitem[Jou96]{Jou96} J. Jourdan, {\sl Nucl. Phys. A} {\bf 603} (1996) 117.
\bibitem[Koh81]{Koh81} M. Kohno and N. Ohtsuka, {\sl  Phys. Lett. B} 
                       {\bf 98} (1981), 335.
\bibitem[Koh83]{Koh83} M. Kohno, {\sl Nucl. Phys. A} {\bf 410} (1983), 349.
\bibitem[Lal97]{Lal97} A.M. Lallena, 
                       {\sl Nucl. Phys. A} {\bf615}, (1997) 325.
\bibitem[Lei90]{Lei90} W. Leidemann and G. Orlandini,
                       {\sl Nucl. Phys. A}{\bf 506} (1990) 447.
\bibitem[Loc75]{Loc75} J.A. Lock and L.L. Foldy, {\sl Ann. Phys.}
                      {\bf 93} (1975), 276. 
\bibitem[Mah82]{Mah82} C. Mahaux and N. Ng\^{o}, {\sl Nucl. Phys. A}
                      {\bf 378} (1982), 205.
                       {\sl Note} CEA-N-2439, Saclay (1985).
\bibitem[Mez84]{Mez84} Z.E. Meziani et al., {\sl Phys. Rev. Lett.} {\bf 52}
                       (1984), 2130.
\bibitem[Mez85]{Mez85} Z.E. Meziani et al., 
                       {\sl Phys. Rev. Lett.} {\bf 54} (1985), 1233; 
                       {\sl DPhN} n$^0$ 2292, Saclay (1984).
\bibitem[Mig57]{Mig57} A.B. Migdal,{\sl Theory of Finite Fermi Systems and
                       Applications to Atomic Nuclei}, Interscience, New
                       York, 1957.
\bibitem[Mon69]{Mon69} E.J. Moniz, 
                       {\sl  Phys. Rev.} {\bf 184} (1969), 1154.
\bibitem[Mon71]{Mon71} E.J. Moniz et al., 
                       {\sl  Phys. Rev. Lett.} {\bf 26} (1971), 445.

\bibitem[Nob81]{Nob81} J.V. Noble, {\sl Phys. Rev. Lett.} {\bf 46} (1981), 412.
\bibitem[Nob83]{Nob83} J.V. Noble, {\sl Phys. Rev. C} {\bf 27} (1983), 423.

\bibitem[Orl91]{Orl91} G. Orlandini and M. Traini, 
                       {\sl Rep. Prog. Phys.} {\bf 54} (1991), 257.

\bibitem[Pec69]{Pec69} R.D. Peccei, {\sl Phys. Rev.} {\bf 181} (1969), 1902.
\bibitem[Pin88a]{Pin88a} D. Pines, K.Q. Quader and J. Wambach,
                        {\sl  Nucl. Phys. A} {\bf 469} (1988), 365
\bibitem[Pin88b]{Pin88b} D. Pines, K.Q. Quader and J. Wambach, 
                       {\sl Nucl. Phys. A} {\bf 477} (1988), 365.
\bibitem[Ris79]{Ris79} D.O. Riska, {\sl en} ``Mesons in nuclei, Vol.II"
                       (M. Rho and D. Wilkinson, Eds.), North-Holland,
                        Amsterdam, 1979.
\bibitem[Ris84]{Ris84} D.O. Riska, {\sl Prog. Part. Nucl. Phys.},
                       {\bf 11} (1984), 199.
\bibitem[Ros50]{Ros50} M.N. Rosenbluth, {\sl Phys. Rev.} {\bf 79} (1950), 615.

\bibitem[Sag89]{Sag89} H. Sagawa, P.M. Boucher, B. Castel and Y. Okuhara,
                      {\sl Phys. Lett. B} {\bf 219} (1989), 10.
\bibitem[Sar93]{Sar93} A.M. Saruis, Phys. Rep. {\bf 235} (1993) 57.
\bibitem[Sch89]{Sch89} R. Schiavilla, V.R. Pandharipande and
                       D.O. Riska, {\sl Phys. Rev. C} {\bf 40}
                       (1989), 2224.  
\bibitem[Shi89]{Shi89} T. Shigehara, K. Shimizu and A. Arima,
                       {\sl Nucl. Phys. A} {\bf 492} (1989), 388.
\bibitem[Sim80]{Sim80} G.G. Simon, Ch. Schmitt, F. Borkowski and
                       V.H. Walther, {\sl Nucl. Phys. A} {\bf 333} (1980), 
                       381.
\bibitem[Sky56]{Sky56} T.H.R. Skyrme,
                       {\sl Phil. Mag.}{\bf 1} (1956) 1043. 





\bibitem[Tra88]{Tra88} M. Traini, S. Turck-Chi\`eze and A. Zghiche
                       {\sl Phys. Rev. C}{\bf 38} (1988), 2799.
\bibitem[Tra93]{Tra93} M. Traini, G. Orlandini and W. Leidemann,
                       {\sl Phys. Rev. C} {\bf 48} (1993) 172. 
\bibitem[Ulm87]{Ulm87} P.E. Ulmer et al,
                       {\sl Phys. Rev. Lett.} {\bf 59} (1987) 2259. 
\bibitem[Vau72]{Vau72} D. Vautherin and D. Brink, 
                      {\sl Phys. Rev. C} {\bf 5} (1972) 626. 
\bibitem[Van81]{Van81} J.W. Van Orden and T.W. Donnelly, {\sl Ann. Phys.
                       (N.Y.)} {\bf 131} (1981), 451.
\bibitem[Van95]{Van95} V. Van der Sluys, J. Ryckebush and M. Waroquier,
                       {\sl Phys. Rev. C} {\bf 51} (1995) 2664.
\bibitem[Voy62]{Voy62} K.W. McVoy and L. Van Hove, 
                      {\sl Phys. Rev.} {\bf 125} (1962) 1034. 
\bibitem[War87]{War87} M. Waroquier et al.,
                       {\sl Phys. Rep.} {\bf 148} (1987) 249.
\bibitem[Weh89]{Weh89} K. Wehrberger and F. Beck, 
                       {\sl Nucl. Phys. A} {\bf 491} (1989), 587.
\bibitem[Wil97]{Wil97} C.F.Wiliamson et al., 
                       {\sl Phys. Rev. C} {\bf 56} (1997) 3152.

\bibitem[Yat93]{Yat93} T.C.Yates et al.,
                       {\sl Phys. Lett.} {\bf B 312} (1993), 382.
\bibitem[Yan82]{Yan82} C. Yannouleas, M. Dworrzecka and J.J. Griffin,
                       {\sl Nucl. Phys. A} {\bf 379} (1982) 256. 
\bibitem[Yan83]{Yan83} C. Yannouleas, M. Dworrzecka and J.J. Griffin,
                       {\sl Nucl. Phys. A} {\bf 397} (1983) 239.
\bibitem[Yen54]{Yen54} D.R. Yennie, D.G. Ravenhall and R.N. Wilson,
                        {\sl Phys. Rev.} {\bf 95} (1954), 500.

\bibitem[Zgh94]{Zgh94} A. Zghiche et al.,
                        {\sl Nucl. Phys. A} {\bf 572} (1994) 513.

 \end{thebibliography}
\end{document}